\documentclass[useAMS,usenatbib,letterpaper]{mnras}
\usepackage{aas_macros,amssymb,graphicx,times,xcolor}
\title[Fully stripped? MACSJ0553$-$3342]{Fully stripped? The dynamics of dark and luminous matter in the massive cluster collision MACSJ0553.4$-$3342} 

\author[]{
 Ebeling H.$^{1}$, Qi J.$^{1}$, Richard J.$^{2}$
\\
$^{1}${Institute for Astronomy, University of Hawaii, 2680 Woodlawn Drive, Honolulu, HI 96822, USA} \\
$^{2}${Univ Lyon, Univ Lyon1, Ens de Lyon, CNRS, Centre de Rech\`erche Astrophysique de Lyon UMR5574, F-69230, Saint-Genis-Laval, France }
}

\date{Submitted to MNRAS on Dec 10, 2016; 2$^{\rm nd}$ revised version}

\pubyear{2017}

\begin{document}
\label{firstpage}
\pagerange{\pageref{firstpage}--\pageref{lastpage}}
\maketitle

\begin{abstract}
We present the results of a multi-wavelength investigation of the very X-ray luminous galaxy cluster MACSJ0553.4$-$3342 ($z{=}0.4270$; hereafter MACSJ0553). Combining high-resolution data obtained with the Hubble Space Telescope and the Chandra X-ray Observatory with groundbased galaxy spectroscopy, our analysis establishes the system unambiguously as a binary, post-collision merger of massive clusters. Key characteristics include perfect alignment of luminous and dark matter for one component, a separation of almost 650 kpc (in projection) between the dark-matter peak of the other subcluster and the second X-ray peak, extremely hot gas (k$T{>}15$ keV) at either end of the merger axis, a potential cold front in the East, an unusually low gas mass fraction of approximately 0.075 for the western component,  a velocity dispersion of $1490_{-130}^{+104}$ km s$^{-1}$, and no indication of significant substructure along the line of sight.

We propose that the MACSJ0553 merger proceeds not in the plane of the sky, but at a large inclination angle, is observed very close to turnaround, and that the eastern X-ray peak is the cool core of the slightly less massive western component that was fully stripped and captured by the eastern subcluster during the collision. If correct, this hypothesis would make MACSJ0553 a superb target for a competitive study of ram-pressure stripping and the collisional behavior of luminous and dark matter during cluster formation.
 \end{abstract}


\section{Introduction}
Galaxy clusters, much more so than the galactic ``island universes'' of Kant's and Herschel's days, are so large and contain so much matter (both dark and luminous) that they effectively constitute mini universes. Serving as self-contained laboratories, massive clusters allow us to study fundamental properties of a wide range of astrophysical objects. This pertains not only to components of the cluster itself (galaxies, gas, and dark matter) but, owing to the magnifying effect of gravitational lensing, also to objects located far behind the cluster \citep[for a review, see][]{kneib11}. In-depth studies of massive galaxy clusters are critical for both of these aspects: observations of the distribution of all three cluster components probe astrophysical processes at the cluster redshift (such as galaxy transformations, structure formation, properties of dark matter, or particle acceleration mechanisms); and the very same observational data also allow us to identify and calibrate these most powerful gravitational lenses for studies of the very distant Universe, as demonstrated by the success of the Hubble Frontier Fields (HFF) project \citep[e.g.,][]{richard14,atek15,bowler15,jauzac15,oesch15}.

Clusters of galaxies that are dynamically disturbed as the result of a recent or ongoing merger are of particular interest. Cluster collisions release an enormous amount of kinetic energy that manifests itself in many forms, including shocks heating the intracluster medium (ICM) and extended radio emission from electrons accelerated to relativistic energies \citep[e.g.,][and references therein]{randall16}. Offsets between the peaks in the gas, galaxy, and dark-matter distribution of merging clusters provide vital clues to the self-interaction cross section of dark matter \citep{markevitch04,clowe06,bradac08,harvey15}. Last, but not least, merging clusters extend over a larger solid angle on the sky and thus offer significant gravitational amplification to a larger number of background galaxies; it is for this reason that five of the six cluster lenses selected for the HFF are merging systems.

We here describe, and present results from, a joint optical / X-ray study of MACSJ0553.4--3342 (MACSJ0553), an X-ray luminous cluster discovered by the Massive Cluster Survey \citep[MACS,][]{ebeling01,ebeling07,ebeling10} and classified as a likely head-on merger by \citet{mann12}. Our paper is organised as follows: having introducing the target of our study in Section~\ref{sec:target}, we provide in Section~\ref{sec:obs} an overview of the X-ray and optical observations used in our work. We then describe in detail our X-ray analysis in Section~\ref{sec:xray} and our approach to constraining the lens model in Section~\ref{sec:sl}. We present our results in Section~\ref{sec:results} and propose a merger scenario that explains all observational evidence in Section~\ref{sec:geom}, before summarising our findings in Section~\ref{sec:summary} and highlighting future research in Section~\ref{sec:future}.

Throughout this paper we adopt the concordance $\Lambda$CDM cosmology, characterised by  $\Omega_{m}=0.3$, $\Omega_{\Lambda}=0.7$, and $H_{0}=70$ km s$^{-1}$ Mpc$^{-1}$. All images are oriented such that north is up and east is to the left.

\section{MACSJ0553.4--3342}
\label{sec:target}

Like all MACS clusters, MACSJ0553 was originally identified as a massive cluster through dedicated optical follow-up imaging of $7{\times}7$ arcmin$^2$ regions around X-ray sources detected in the course of the ROSAT All-Sky Survey \citep[RASS,][]{voges99,boller16}; a detailed description of the MACS project can be found in \citet{ebeling01}.  Images in the V, R, and I bands obtained in February 2002 with the University of Hawai'i's 2.2m telescope (UH2.2m) on Maunakea unambiguously established the counterpart to the RASS source 1RXS\,\,J055326.7--334237 as an optically rich galaxy cluster. A preliminary cluster redshift of $z{=}0.431$ was established by long-slit spectroscopy conducted with the Keck-II Echelle Spectrograph and Imager (ESI), leading to a first estimate of the X-ray luminosity of the system of $L_{\rm X}=(1.63\pm0.25){\times}10^{45}$ erg s$^{-1}$ (0.1--2.4 keV) derived from the 57 photons detected from 1RXS\,\,J055326.7--334237 in the RASS.

\begin{figure}
\hspace*{-5mm}\includegraphics[width=0.5\textwidth]{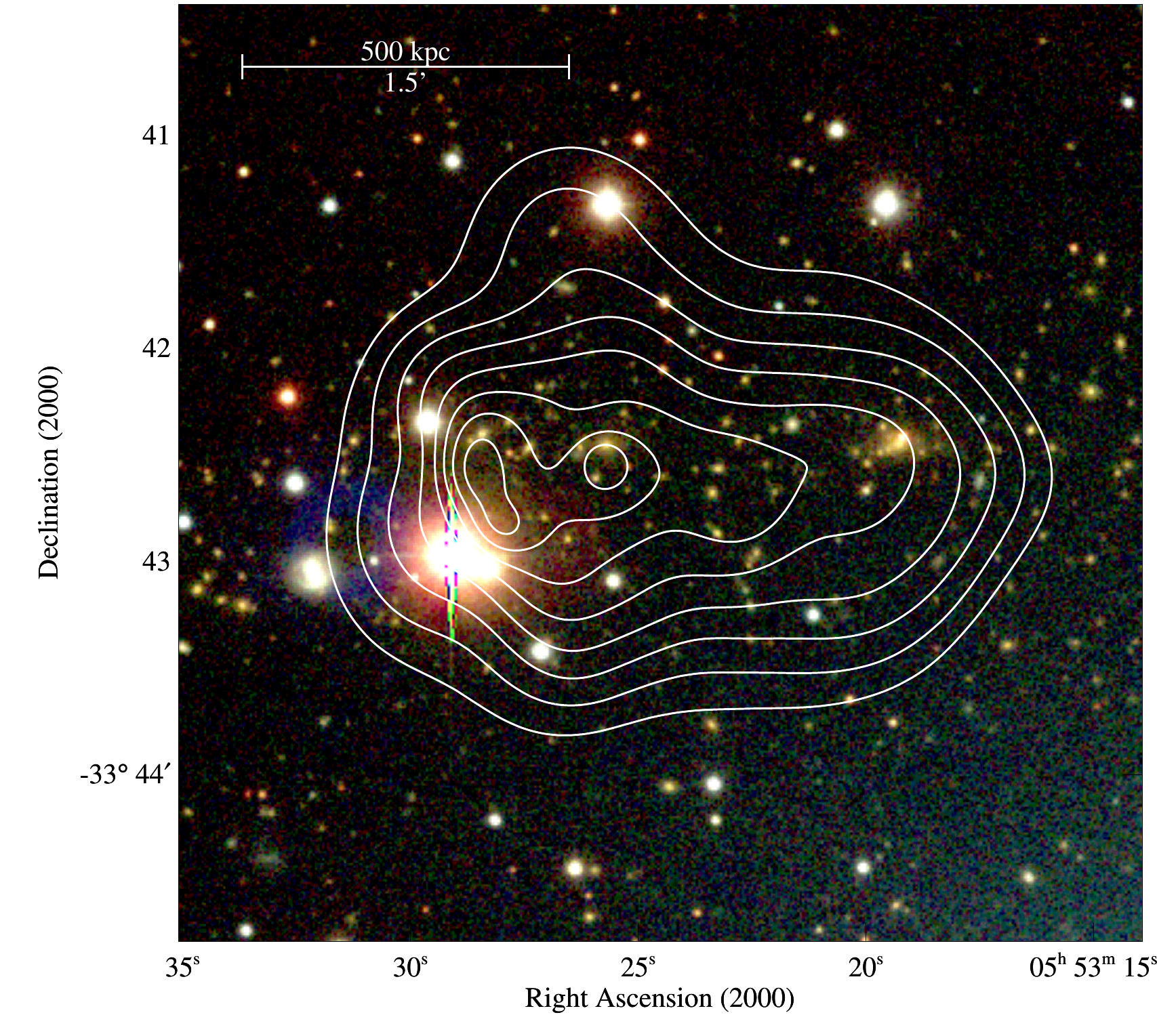}
\caption{Iso-intensity contours of the adaptively smoothed X-ray emission from a 10ks Chandra observation of MACSJ0553 overlaid on a colour-composite image created from UH2.2m observations conducted in February 2011 in the V, R, and I filters. The distribution of both galaxies and intra-cluster gas is far from regular and relaxed. Reproduced from \citet{mann12}.  \label{fig:uh88}}
\end{figure}

Based on the galaxy distribution shown in deeper UH2.2m images (taken in February 2011) and the ICM morphology revealed by a short Chandra X-ray observation of the cluster (Fig.~\ref{fig:uh88}), \citet{mann12} classified MACSJ0553 as a likely head-on merger and a promising target for an improved measurement of the dark-matter self-interaction cross section. Motivated by this assessment, we successfully proposed a joint Chandra/HST in-depth study of the system in Chandra Cycle 12 (HST Cycle 18). The observations performed for this project are briefly summarised in the following section.

\section{Observations and Data Reduction}
\label{sec:obs}

\subsection{Optical}

\subsubsection{HST Imaging}
MACSJ0553 was first observed with the Hubble Space Telescope (HST) in 2008, using the Wide-Field and Planetary Camera 2 \citep[WFCP2;][]{trauger94} for our MACS SNAPshot program GO-11103 (PI: Ebeling). The resulting image in the F606W passband (total exposure time: 1200s) is, however, not used by us here, due to the availability of far superior data from subsequent HST observations.
 
\begin{figure}
\hspace*{-0mm}\includegraphics[width=0.5\textwidth,clip=true,trim=110 5 110 5]{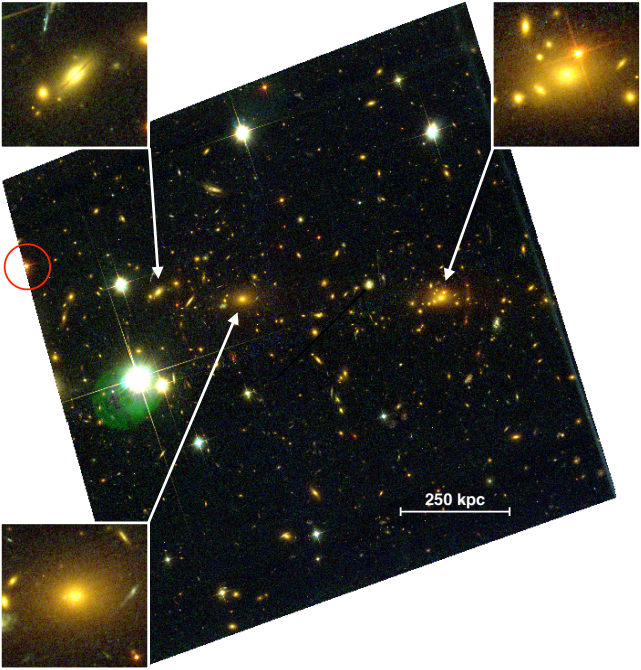}
\caption{MACSJ0553 as seen with HST/ACS (F435W/F606W/F814W). A comparison with Fig.~\ref{fig:uh88} shows that our ACS observation covers essentially all of the area within with X-ray emission was detected with Chandra. Insets ($15\arcsec{\times}15\arcsec$) show the brightest cluster galaxies (BCGs); note that the third-brightest galaxy (top left) is a foreground object (see Section~\ref{sec:gals}). The projected distance between the BCGs corresponds to 446 kpc at the cluster redshift. The red circle at the eastern edge of the field marks a binary star used to ensure astrometric alignment with the X-ray data (see Section~\ref{sec:xray}).\label{fig:hst}
}
\end{figure}

Observations with HST's Advanced Camera for Surveys \citep[ACS;][]{ford98} in the F435W, F606W, and F814W filters were executed for GO-12362 (PI: Ebeling) in January 2012 for total exposure times of 4452, 2092, and 4572s, respectively. Additional shallow coverage at near-infrared wavelengths was added through short exposures (656s and 381s) with the Wide Field Camera 3 \citep[WFC3;][]{kimble08} in November 2016 for program GO-14096 (PI: Coe) in the F105W and F140W filters. We used the \textsc{astrodrizzle} v2.1.0 software package to remove geometric distortions, background variations, and cosmic rays, and to create combined images with a pixel size of 0.03\arcsec (we set \texttt{pixfrac}=0.7 as recommended by ISR ACS 2015-1). The resulting ACS colour image (Fig.~\ref{fig:hst}) covers approximately the same area as shown in Fig.~\ref{fig:uh88} and confirms the bimodal nature of the system already apparent in the groundbased image, anchored by two approximately equally luminous brightest cluster galaxies (BCGs), separated by  almost 450 kpc in projection.

\subsubsection{Spectroscopy}
\label{sec:spec}

After the redshift of MACSJ0553 was initially established in Keck-II/ESI observations, the system was targeted four more times by us in multi-object spectroscopy observations with the LRIS and DEIMOS spectrographs on Keck-I and -II, respectively (see Table~\ref{tab:obsruns} for a summary). A slit width of 1\arcsec was chosen for all observations. 

With LRIS, we used the D6800 dichroic to split the incoming light into a blue and red part, directed at the respective arm of the spectrograph. For the red arm, the 600/7500 grating, set to a central wavelength of 8100\AA, yielded an effective resolution of 5\AA; on the blue side the 300/5000 grism provided 9\AA\ resolution. With DEIMOS, we used the 600 l/mm grism and the G455 blocking filter to suppress second-order contamination at wavelengths exceeding 9100\AA; the effective spectral resolution achieved in these observations is 3.5\AA. 

We used the XIDL and DEEP2 \citep{cooper12,newman13} pipelines to reduce the LRIS and DEIMOS data, respectively. Redshifts were measured from the one-dimensional spectra using elements of the SpecPro software package \citep{masters11}.

\begin{table}
\begin{tabular}{lccc}
Instrument & Date & exposure time & seeing\\ \hline \\
LRIS & Dec 2011 & blue arm: 3600s & 1.2--1.4\arcsec\\
         &                 & red arm: 3240s  &\\
DEIMOS & Nov 2014 & mask 1: 1600s & 0.9\arcsec\\
               &                 & mask 2: 3000s & 0.8\arcsec\\
DEIMOS & Nov 2015 & mask 3: 4931s & 1.5\arcsec\\
DEIMOS & Oct 2016 & mask 4: 3300s & 0.9\arcsec\\
               &                 & mask 5: 2400s & 1.0\arcsec\\
\end{tabular}
\caption{Overview of our spectroscopic observations of MACSJ0553 conducted on Keck I and II. When two masks were observed in the same observing run, all slits on strong-lensing features were common to both masks, resulting in exposure times of 4600s and 5700 for Nov 2014 and Oct 2016, respectively. The listed seeing is the median-averaged FWHM of the spatial profile for alignments stars. \label{tab:obsruns}}
\end{table}

\subsection{X-ray}
\label{sec:xover}
MACSJ0553 was observed with the Chandra X-ray Observatory's ACIS-I detector in 2005 (ObsID 5813: PI Ebeling) and again in 2011 (ObsID 12244, PI: Ebeling) for exposure times of 9.9 and 71.9 ks, respectively\footnote{We quote the exposure times of either observation after filtering out periods of high background (flares).}. Both observations were performed in Very Faint mode. We reprocessed the datasets from either observation with the standard CIAO analysis tools and Chandra Calibration Database 4.7.2. Periods of flaring background were identified and removed with the help of the \texttt{lc\_clean} script, before projection of the shorter observation onto the longer one.

\begin{figure}
\hspace*{-0mm}\includegraphics[width=0.5\textwidth]{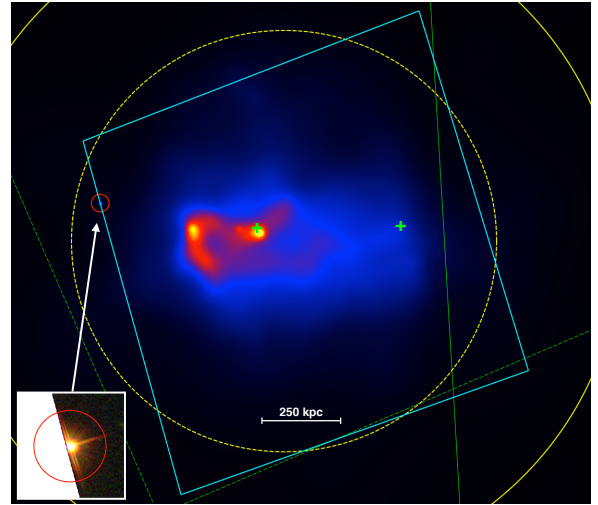}
\caption{MACSJ0553 as seen with Chandra/ACIS-I in our merged 82 ks observation; the data were adaptively smoothed to $3\sigma$ significance. The green crosses mark the location of the BCGs highlighted in Fig.~\ref{fig:hst}; $r_{\rm 2500}$ and $r_{\rm 1000}$ are marked by the dashed and solid yellow circle, respectively. The field of view covered by our HST/ACS observation is outlined in cyan, and the ACIS-I chip gaps for ObsIDs 5813 and 12244 are indicated by dashed and solid green lines, respectively.  Finally, the inset ($15\arcsec{\times}15\arcsec$) shows the perfect alignment between a faint X-ray point source and its optical counterpart, a binary star (cf.\ Fig~\ref{fig:hst}).\label{fig:cxo}
}
\end{figure}

Fig.~\ref{fig:cxo} shows the X-ray emission from MACSJ0553 in the Chandra broad band (0.5--7 keV) after adaptive smoothing of the merged dataset. We use \textsc{asmooth} \citep{ebeling06} with the default setting of $3\sigma$ for the significance with respect to the local background of any features in the adaptively smoothed image. Note the highly irregular morphology of the X-ray emission and the total absence of an X-ray peak within over 400 kpc of the BCG of the western subcluster.

We take advantage of an X-ray binary within the ACS field of view, detected as a faint point source in our deep ACIS-I observation, to check the consistency of the astrometric solutions of the HST and Chandra observations. We find only a very small offset of 0.44\arcsec for which we adjust, using the HST astrometry as our reference. For the characterisation of the distribution and temperature of the ICM we then mask out all point sources detected by the {\tt celldetect} tool at $3\sigma$ significance. The subsequent analysis of the resulting cleaned X-ray data, including the treatment of background emission, is described in detail in Section~\ref{sec:xray}.

\subsection{Radio}

\begin{figure}
\hspace*{-0mm}\includegraphics[width=0.5\textwidth]{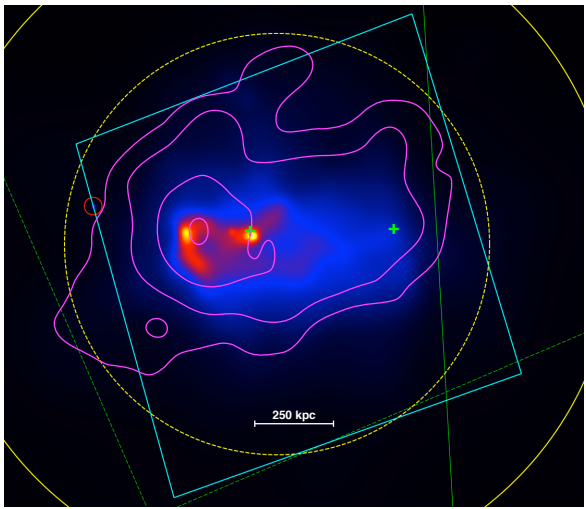}
\caption{Like Fig.~\ref{fig:cxo}, but showing overlaid in magenta the radio contours from GMRT observations at 323 GHz, reproduced from \citet{bonafede12}. \label{fig:radio}
}
\end{figure}

MACSJ0553 was observed with the Giant Metrewave Radio Telescope (GMRT) at 323 GHz by \citet{bonafede12}  who detected a very extended radio halo but no sign of the radio relics they expected from a merger proceeding in the plane of the sky. The GMRT contours are shown in Fig.~\ref{fig:radio}, overlaid on our X-ray image of MACSJ0553.

\section{X-ray Analysis}
\label{sec:xray}

\subsection{Background subtraction}
\label{sec:xbkg}

Background emission recorded in X-ray observations can be divided into two components of different physical origin: photons and high-energy particles. The former component is a combination of unresolved emission from faint sources at high redshift and diffuse emission from gas in our own Galaxy; as such, this component is temporally stable. The latter component, however, which by far dominates the background seen with Chandra's ACIS instrument, is time variable, primarily due to the variability of the Sun's activity. 

While it may thus seem advisable to determine the background from the science observation itself by selecting apparently source-free regions of the detector, this strategy has several drawbacks. For one, such designated background regions will, by necessity, fall onto a different part of the detector than the science target and are thus subject to a different instrumental response. Also, the area that can be used to determine the background is restricted to regions well away from the target; if the X-ray emission from the latter is significantly extended, the size of the background region is necessarily reduced. Finally, the accuracy of the background measurement is limited by the statistics set by the duration of the science exposure.

\subsubsection{Blank-sky background}

To avoid these complications we here follow the alternative path of determining the X-ray background from separate calibration files, provided by the CALDB for the time of our observation via the \texttt{acis\_bkgrnd\_lookup} tool. As a first step, we project the blank-sky and particle-background\footnote{These observations were performed with ACIS  stowed inside the detector housing and thus do not contain any photon events; by contrast, the very deep blank-sky observations contain both photon and particle background.} datasets thus obtained to the pointing coordinates and orientation of each of our two Chandra observations of MACSJ0553. Following \citet{hickox06} we then account for the time variability of the particle background by scaling the archival particle background such that it matches our observations in the 9--12 keV range. The required scaling factor is calculated as:
\begin{equation}
		A = \frac{c({\rm 9{-}12\, keV}) - 0.0132 \times c_{\rm readout}({\rm 9{-}12\, keV})}{c_{\rm particle}({\rm 9{-}12\, keV})} \label{eqn:scale}
\end{equation}
where $c$ and $c_{\rm particle}$ are the count rates in the 9--12 keV range recorded in the photon events files for a given ACIS observation and for the time-matched archival particle-background observation, respectively, while $c_{\rm readout}$ is the readout contribution to the background, provided by the CIAO tool \texttt{readout\_bkg}. Properly scaled, the particle background $A\times f_{\rm particle}$ can then be subtracted across the full spectrum within any given region.

Since our analysis works with fluxes, rather than counts, we need to ensure that differences in exposure time as well as the telescope and instrument response are accounted for. We use the CIAO \texttt{specextract} tool to extract spectra from our science data set and to compute the corresponding auxiliary response and redistribution matrix files (ARF and RMF). For the blank-sky and particle-background data sets ARFs and RMFs are not available. We therefore apply the respective ARF and RMF from the science observation to blank-sky and particle-background spectra generated with \texttt{dmextract}, and then scale both by appropriate factors to match the exposure time of the science observation.
 
 We implement the subtraction of the background by combining two techniques. We directly subtract the appropriately scaled particle background, both from our science observation and the blank-sky dataset, within the \textsc{sherpa} environment provided as part of the CIAO software package. A different approach is taken for the photon background and described in the following section.
 
\subsubsection{Photon background model}

 The photon background is not subtracted, but modeled as the sum of thermal emission from the Galactic bubble \citep{kuntz00,lumb02} and a power-law contribution that accounts for unresolved emission from distant AGN. We add a multiplicative exponential term to account for photoelectric absorption by the Galaxy. This model is then fit to the blank-sky data set with all model parameters frozen at their literature values except for their normalisations and the equivalent column density of neutral hydrogen which we set to the value determined from 21-cm observations in the direction of each target \citep{kalberla05}. By fitting both blank-sky data sets simultaneously (one each for ObsIDs 5813 and 12244) and requiring their respective normalizations to match the ratio of the exposure times, we obtain a universal, fully determined model for the photon background that, properly scaled to the area under consideration, can be included in the spectral fits to the data for any region of our science dataset. The parameters defining this photon background model are listed in Table~\ref{tab:xbg}.

\begin{table}
\begin{tabular}{ll}
Component & CIAO model and parameter values\\ \hline
Black Body I: & \textsc{xsapec}($N$, k$T$=0.14 keV, $z$=0, $Z{=}Z_\odot$)\\
Black Body II: & \textsc{xsapec}(0.3164${\times}N$, k$T$=0.248 keV, $z$=0, $Z{=}Z_\odot$) \\
Power Law: & \textsc{powlaw1d(0.3551${\times}N$, $\Gamma$=1.42)}\\
\end{tabular}
\caption{Summary of the components of our model for the photon background as provided by  \citet{kuntz00} and \citet{lumb02}. The only free parameter, $N$, sets the overall normalization of the photon background; we find $N{=}1.986{\times}10^{-4}$ from a fit to the blank-sky spectra for the I3 CCD of ACIS-I (merged data set). \label{tab:xbg}}
\end{table}

\subsection{Spectral model}

Prior to fitting, spectra from both data sets (ObsIDs 5183 and 12244) are combined using \texttt{specextract}; the same procedure is applied to the corresponding particle-background data sets. We then group the spectra, requiring 20 counts per bin, except for fits to the global emission (see Section~\ref{sec:icm}) where we require 30 counts per bin. After subtraction of the particle-background component we fit a thermal-plasma model for the cluster emission (plus a fixed photon-background contribution) modified by absorption at the observer. In \textsc{sherpa} terms, our model is thus described as \texttt{xswabs}$\times$ (\texttt{xsapec.sc}+\texttt{xsapec.bg1}+\texttt{xsapec.bg2}+\texttt{powlaw1d.bg3}), where the final three components describe the photon background (see Table~\ref{tab:xbg}) and are scaled to the area of the region of interest.

\subsection{Temperature Trend Maps}
\label{sec:ttrend}

In order to investigate and quantify spatial variations of the ICM temperature in MACSJ0553, we employ the adaptive-grid algorithm \textsc{contbin} \citep{sanders06} to create an irregular tessellation whose cells all share the same, pre-set signal-to-noise ratio (SNR) for the net photon counts as recorded in the Chandra broad-band image (0.5--7 keV) of MACSJ0553. By design, \textsc{contbin}  generates a set of regions that provide high-resolution sampling in areas of high X-ray flux, and coarse sampling where the signal is close to the background. Note that the constant-SNR criterion can be met by a  large number of different tessellations; the solution returned by \textsc{contbin} is just one of many. This can cause significant systematic effects if the quantity computed within each cell is in fact not correlated with flux -- one example of such an uncorrelated quantity being the ICM temperature. In the following we briefly describe a method developed by us that effectively bootstraps the uncertainties introduced by the precise choice of a cell's geometric shape and placement within a region of a given SNR.

\begin{figure}
\hspace*{-0mm}\includegraphics[width=0.49\textwidth]{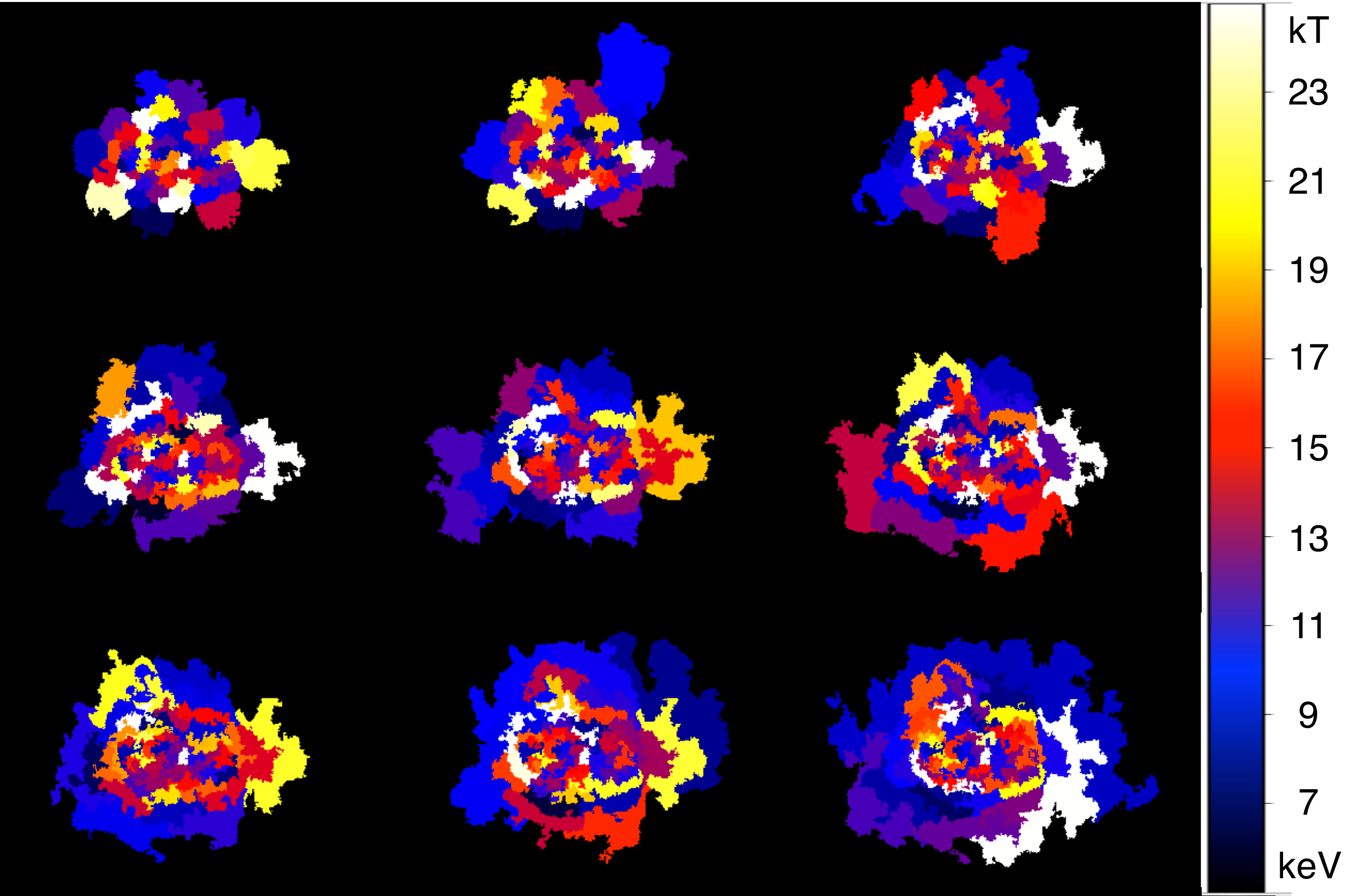}\\[3mm]
\hspace*{-0mm}\includegraphics[width=0.49\textwidth]{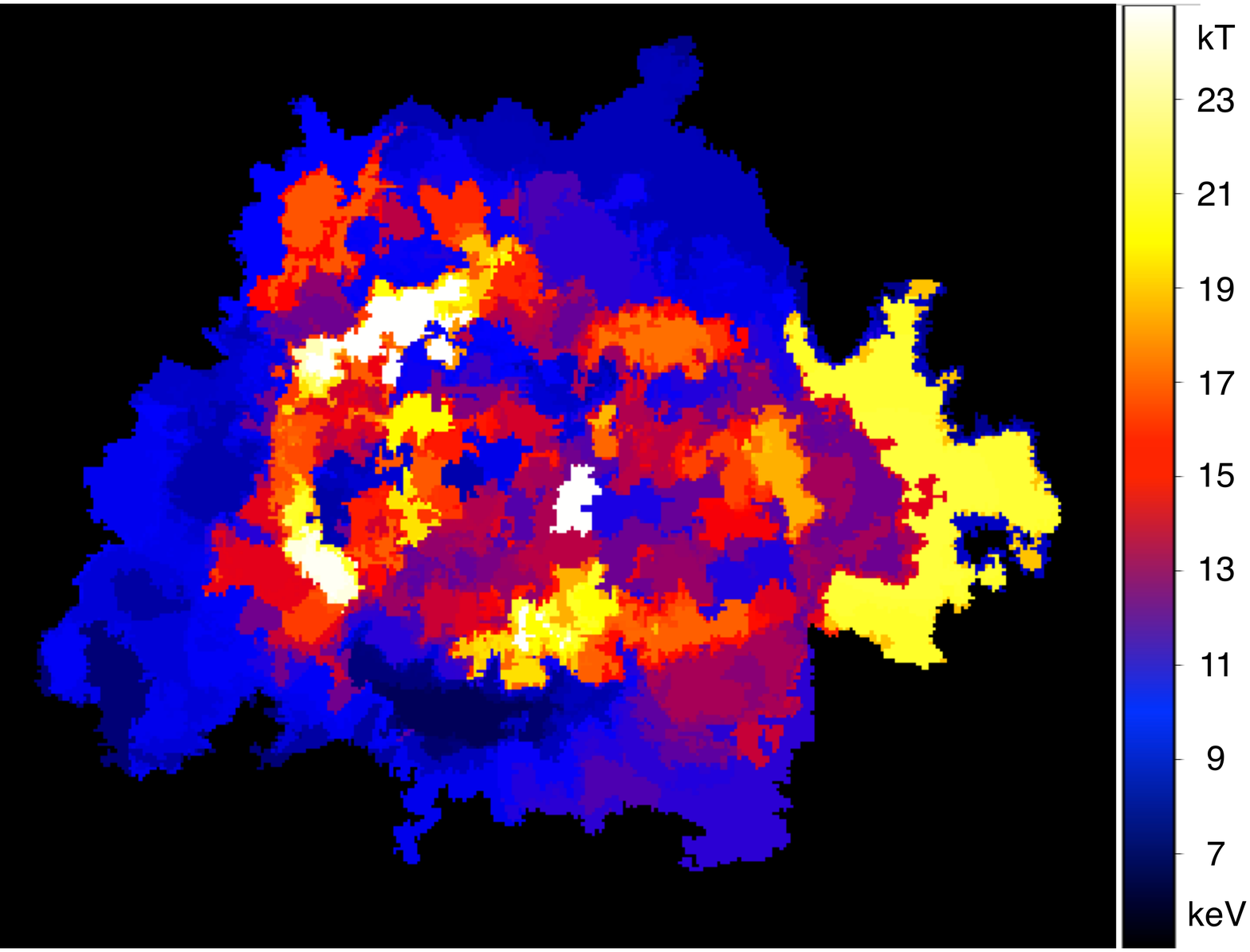}
\caption{Top: temperature maps created with \textsc{contbin} for SNR=20 and maximal ellipticities of 1.1, 1.2,  ... 1.9 (left to right, and top to bottom); cells touching the edges of our data region, and thus dominated by background emission, have been removed. Bottom: temperature-trend map for SNR=20, created by median averaging the pixel values from the nine individual temperature maps shown in the top panel. \label{fig:kttrends}
}
\end{figure}

In order to sample the distribution of possible tessellations and reduce ``geometric bias", we compute temperature maps for nine variations of the default \textsc{contbin} tessellation, created by varying \texttt{constrainval} (which specifies the upper limit on the ellipticity of any given cell) from 1.2 to 2.0 in steps of 0.1. The resulting array of temperature maps\footnote{We remove tiles that touch the edge of our data region in order to mitigate the impact of areas dominated by background.} for an SNR value of 20 is shown in Fig.~\ref{fig:kttrends} (top). We then median-average the pixel values in this stack of nine images to arrive at the map shown in the lower panel of Fig.~\ref{fig:kttrends} which suppresses the geometric bias present in any given tessellation and enhances robust temperature trends that are common to a plurality of realisations. We stress that the apparently much higher resolution provided by the median image is merely the result of the applied pixel-by-pixel averaging process; in general, none of the apparent cells in the median image comes close to meeting the SNR criterion set by us for the counts in the Chandra broad band. To avoid misunderstandings, we thus refer to these geometrically randomised images as \textit{Temperature Trend Maps}.

\subsection{Spatial modelling}
\label{sec:beta2d}

Although the heavily disturbed X-ray morphology of MACSJ0553 (individual features are discussed in Section~\ref{sec:xsb}) and the limited photon statistics of the existing Chandra observations prevent a physically meaningful modeling of the observed surface brightness distribution (let alone any deprojection analysis), we nonetheless attempt to quantify and compare the large-scale X-ray emission from the ICM of the two apparent cluster components anchored by the system's two BCGs. To this end we bin the X-ray data within a $6.3\times6.3$ arcmin$^2$ region (approximately 1 Mpc on the side) into 3\arcsec\ pixels, and then remove the irregularly shaped X-ray peak in the far East of the system (see Fig.~\ref{fig:cxo}) by replacing the pixel values in this area with the azimuthal average observed on either side, adopting the western X-ray peak as the centre, and adding noise commensurate with the photon statistics. We subsequently fit two elliptical, two-dimensional $\beta$ models \citep[][\textsc{beta2d} in \textsc{sherpa}]{beta76} to the resulting surface-brightness image,
\[S(\hat{r})=S_0 \left[1+\left({\hat{r} \over r_0}\right)^2\right]^{-3\beta+0.5}, \]
where $\hat{r}$ is the projected, elliptical radius on the sky, $S_0$ is the central surface brightness, $\beta$ is the power index, and $r_0$ is the  core radius (see \url{http://cxc.harvard.edu/sherpa/ahelp/beta2d.html} for details). The gas density is then given by 
\[\rho(r)=\rho_0 \left[1+\left({r \over r_0}\right)^2\right]^{-{3\over2}\beta},\]
with $r$ and $\rho_0$ being the three-dimensional radius and the central gas density, respectively. $\rho_0$ is obtained by equating the integral of the emission measure to the normalization of the best-fitting spectral model for the emission within the same spatial region, as described in, e.g., \citet{ho12} or \citet{david12}. 

Although the $\beta$ model usually provides a satisfactory description of the X-ray surface brightness distribution of isothermal, relaxed clusters, its single power-law slope is unable to reproduce the radial steepening of the X-ray emission frequently observed at large radii for non-relaxed systems. In addition, its integral, i.e., the total gas mass, diverges for shallow profiles, i.e., low values of $\beta$. For both of these reasons, an additional, multiplicative term,
\[
\left[1+\left({r\over r_s}\right)^\gamma\right]^{-{\epsilon\over\gamma}}
\]
was proposed by \citet{vikhlinin06} to reproduce any steepening in the model's radial dependence at a large radius $r_s$. While \citeauthor{vikhlinin06} find a value of 3 for $\gamma$ (the parameter controlling the width of the transition region between the two power-law slopes) to be a universally acceptable choice, their values for both $r_s$ and $\epsilon$ vary dramatically between clusters, and neither parameter is constrained within the region where our only modestly deep Chandra observation is able to detect the ICM's diffuse X-ray emission (Fig.~\ref{fig:cxo}). We explore the resulting systematic errors by computing cumulative gas-mass profiles for two plausible values for each of $r_s$ and $\epsilon$ (1000 and 1500 kpc, and 2 and 3, respectively) for both components of our double $\beta$ model.

\section{Strong-lensing Analysis}
\label{sec:sl}
In the strong-lensing regime, the gravity-induced deflection of light emitted by a distant background object located near a caustic in the source plane can create multiple images of the object. Since gravitational lensing is achromatic and conserves surface brightness, such multiple images can often be readily identified by their matching colours and morphologies, enabling us to invert the lens equation to derive the mass distribution within the cluster lens \citep[e.g.,][]{mellier93,kneib96,kneib11}.

Our approach to modeling the mass distribution of MACSJ0553 using such strong-lensing constraints follows the strategy applied successfully before by us, most recently for the calibration of the massive cluster lenses selected for the Hubble Frontier Fields project \citep[e.g.,][]{richard14}. Specifically, we use \textsc{Lenstool} \citep{jullo07} to optimise a parametric model of the mass distribution of the cluster, built from both cluster-size large-scale halos and smaller, galaxy-sized perturbers, in order to reproduce the strong-lensing constraints derived from the data. In doing so, we adopt dual pseudo-isothermal elliptical mass distributions for the model components \citep{richard09,limousin12} which are characterised by seven parameters (right ascension and declination of the centre, normalisation, core radius, cutoff radius, ellipticity, and orientation).

\begin{figure*}
\hspace*{-0mm}\includegraphics[width=\textwidth]{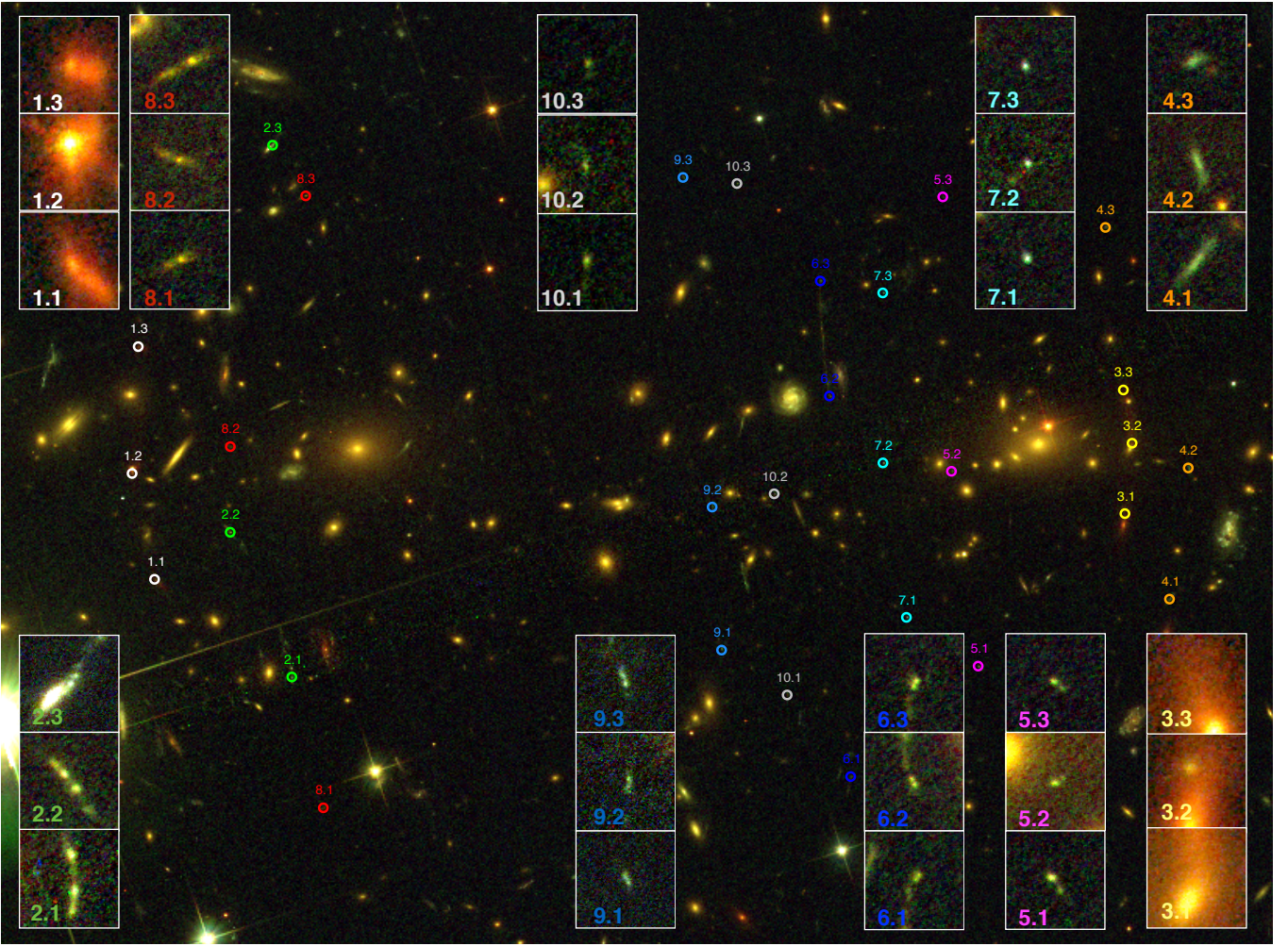}
\caption{As Fig.~\ref{fig:hst}, but highlighting strong-lensing features. Ten multiple-image systems are marked; insets show a $3{\times}3$ arcsec$^2$ area around the individual images for each family at a more aggressive stretch. For systems 1 and 3, the reddest multiple-image families, we show postage stamps generated from F435W/F814W/F140W images.\label{fig:hstsl}
}
\end{figure*}

Cluster members acting as galaxy-scale perturbers are identified by their colours where spectroscopic redshifts (see Section~\ref{sec:spec}) are not available. We adopt values for the location, ellipticity, and orientation as measured by \textsc{SExtractor} \citep{bertin96} from the galaxies' optical appearance in the ACS/F814W image. All other parameters of galaxy-scale mass components are set based on the assumption of a constant mass-to-light ratio: specifically the central velocity dispersion, core and cut-off radii are scaled according to their luminosity, using the respective values of $\sigma_0^*$, $r_{\rm core}^*$ and $r_{\rm cut}^*$ for an $L^*$ galaxy. Since $r_{\rm core}^*$ is small, we follow the approach taken by \citet{richard10} for  a similar strong-lensing analysis of clusters at $z{<}0.3$ by adopting a fixed value $r_{\rm core}^*{=}0.15$ kpc and optimising the values of $\sigma_0^*$ and $r_{\rm cut}^*$ using a Gaussian prior $\sigma_0^*=(158\pm27)$ km/s and a flat prior in the range $10{<}r_{\rm cut}^*{<}100$ kpc, respectively.

\subsection{Strong-lensing features}

Our search for strong-lensing features starts from the most obvious sets of multiple images apparent in the HST colour image of MACSJ0553, the bright arcs highly elongated in the north-south direction (labeled systems 3 and 6 in Fig.~\ref{fig:hstsl}). We then iteratively refine the model by adding further constraints imposed by multiple-image systems identified based on colour and morphology as well as redshift information. Figure~\ref{fig:hstsl} shows an annotated version of the HST/ACS colour image of MACSJ0553 where we highlight ten multiple-image families identified in this manner; their location and redshifts (either measured or estimated) are listed in Table~\ref{tab:hstsl}.

\begin{figure}
\hspace*{0mm}\includegraphics[width=0.48\textwidth]{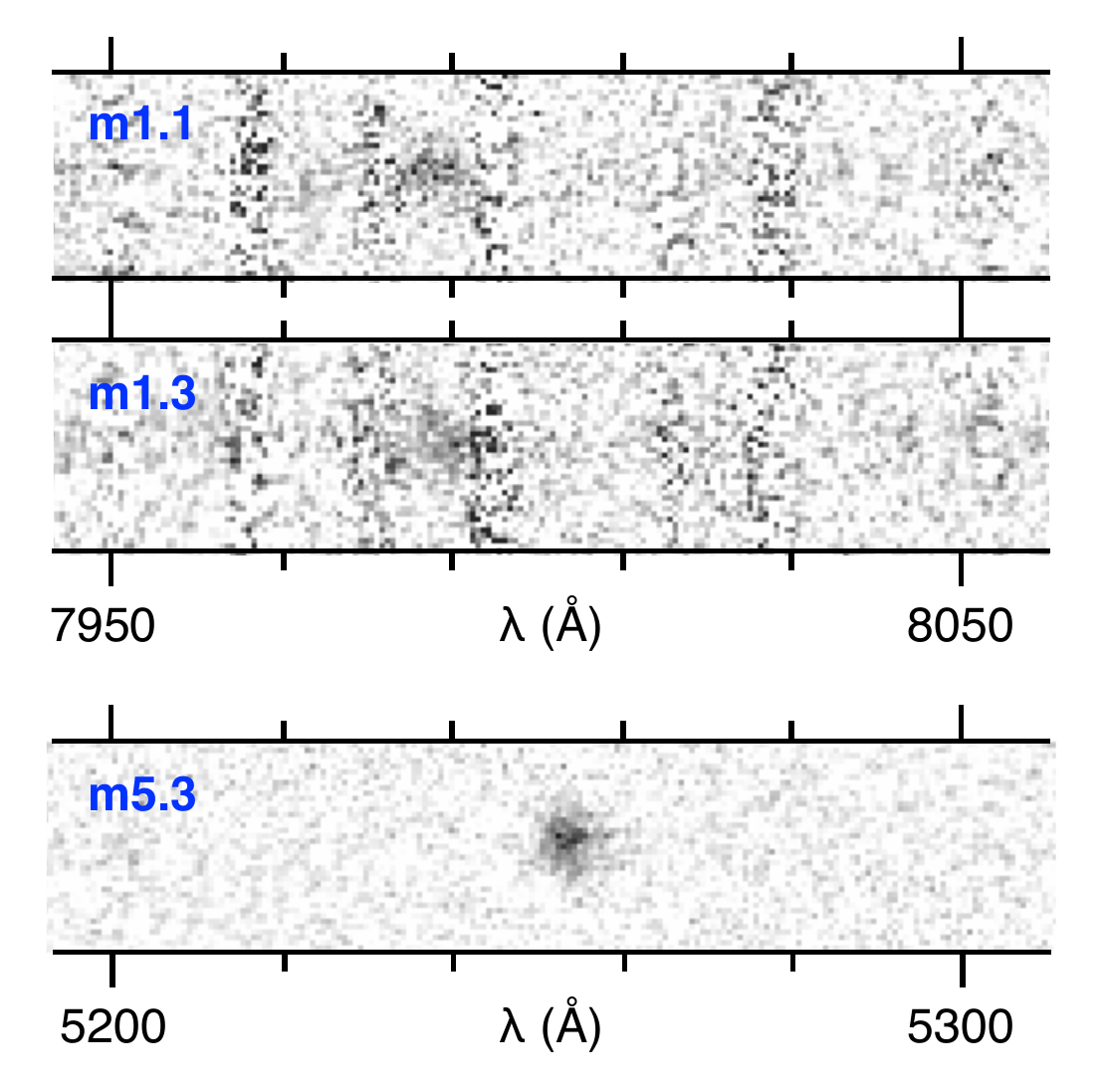}
\caption{Part of the spectra obtained by us with Keck-II/DEIMOS for images m1.1, m1.3, and m5.3. We show the features identified as [{O\,\textsc{\lowercase{ii}}}]$\lambda$3727 (in multiple-image system m1) and Ly$\alpha$ (in multiple-image system m5), respectively. \label{fig:spec2d}
}
\end{figure}

Spectroscopic redshifts were obtained with Keck-II/DEIMOS (see Section \ref{sec:spec}) for system 1 (where we measure the redshifts for images 1.1 and 1.3 from the detection of the [{O\,\textsc{\lowercase{ii}}}] emission line), and for system 5 (from the identification of strong Lyman-$\alpha$ emission for image 5.3). We show the relevant feature from the spectra of these three images in Fig.~\ref{fig:spec2d}. Our spectroscopic observations also targeted all images of system 4 and 7, as well as the multiple-image family members 2.1, 2.2, 8.1, and 8.3 without detecting spectral features in any of them. For systems 2, 3, 4, 7, 8, 9, and 10, we therefore use the ACS photometry in three bands to estimate a crude photometric redshift with the \textsc{HyperZ} code \citep{bolzonella00}; the results then refine the redshift prior for the Lenstool model as follows. We adjust the spectral energy distribution of each multiple image with spectral templates at solar metallicity \citep{BC03}, varying the reddening within $0<A_V<4$ and the redshift within $0.5<z<7$; we then use the resulting redshift probability distribution $P(z)$ from each system as a prior for the model (typically a Gaussian prior centred on the best $z$ and with $\sigma_z{=}$0.5--1.2). The actual redshift of each system is optimised as a parameter in Lenstool.

\subsection{Model optimisation}

The optimisation of our parametric strong-lensing model of the mass distribution with Lenstool is performed through a Bayesian Markov Chain Monte Carlo (MCMC) process, in which all model parameters are varied and a likelihood function is computed based on how well the model reproduces the locations of all multiple-image systems in the image plane\footnote{We adopt a value $\sigma_{\rm pos}=0.5\arcsec$ for the systematic uncertainty in the position of each image.}.

We use the results of the MCMC sampling of the posterior probability distributions to derive the best fit and relative errors on all model parameters. 

\begin{table}
\begin{tabular}{lccccc}
ID & R.A.\ (J2000) & Dec. (J2000) & $z_{\rm prior}$ & $z_{\rm model}$ & $z_{\rm spec}$\\ \hline
1.1 & 05:53:27.65 & $-$33:42:43.1 & & & 1.14 \\ 
1.2 & 05:53:27.86 & $-$33:42:30.6 & & & \\ 
1.3 & 05:53:27.80 & $-$33:42:15.6 & & & 1.14 \\ \hline
2.1 & 05:53:26.36 & $-$33:42:54.7 & [1.5--3.9] & 3.11$\pm$0.14 & \\ 
2.2 & 05:53:26.94 & $-$33:42:37.6 & & & \\ 
2.3 & 05:53:26.54 & $-$33:41:51.7 & & & \\ \hline
3.1 & 05:53:18.53 & $-$33:42:35.4 & [0.4--2.4] & 0.97$\pm$0.06 & \\ 
3.2 & 05:53:18.46 & $-$33:42:27.0 & & & \\ 
3.3 & 05:53:18.55 & $-$33:42:20.7 & & & \\ \hline
4.1 & 05:53:18.11 & $-$33:42:45.5 & [1.4--3.8] & 2.83$\pm$0.19 & \\ 
4.2 & 05:53:17.94 & $-$33:42:30.0 & & & \\ 
4.3 & 05:53:18.72 & $-$33:42:01.5 & & & \\ \hline
5.1 & 05:53:19.91 & $-$33:42:53.4 & & & \\ 
5.2 & 05:53:20.16 & $-$33:42:30.4 & & & \\ 
5.3 & 05:53:20.24 & $-$33:41:57.9 & & & 3.32\\ \hline
6.1 & 05:53:21.11 & $-$33:43:06.5 & [3.1--4.2] & 4.03$\pm$0.18 &\\    
6.2 & 05:53:21.31 & $-$33:42:21.4 & & & \\     
6.3 & 05:53:21.39 & $-$33:42:07.8 & & &\\ \hline
7.1 & 05:53:20.58 & $-$33:42:47.7 & [0.4--3.0] &  1.48$\pm$0.05 &\\ 
7.2 & 05:53:20.81 & $-$33:42:29.4 & & &\\ 
7.3 & 05:53:20.81 & $-$33:42:09.2 & & &\\ \hline
8.1 & 05:53:26.06 & $-$33:43:10.2 & [2.5--6.0] & 5.39$\pm$0.35 &\\ 
8.2 & 05:53:26.94 & $-$33:42:27.4 & & &\\ 
8.3 & 05:53:26.23 & $-$33:41:57.7 & & &\\ \hline
9.1 & 05:53:22.32 & $-$33:42:51.5 & [1.8--4.2] & 2.62$\pm$0.12 &\\ 
9.2 & 05:53:22.41 & $-$33:42:34.6 & & &\\ 
9.3 & 05:53:22.68 & $-$33:41:55.6 & & &\\ \hline
10.1 & 05:53:21.71 & $-$33:42:56.8 & [3.3--4.3] & 3.31$\pm$0.14\ &\\ 
10.2 & 05:53:21.83 & $-$33:42:33.0 & & &\\ 
10.3 & 05:53:22.18 & $-$33:41:56.3 & & &\\ 
\end{tabular}
\caption{Multiple-image systems as marked in Fig.~\ref{fig:hstsl}. Spectroscopic redshifts are listed for the image for which they were measured, but are set as priors for all images of the same family. \label{tab:hstsl}}
\end{table}

\section{Results}
\label{sec:results}

\begin{table}
\begin{tabular}{ll}
Overdensity radii: &$r_{\rm 200}{=}2.32$ Mpc\\
                                &$r_{\rm 1000}{=}1.04$ Mpc\\
                                &$r_{\rm 2500}{=}657$ kpc\\ \hline
Systemic redshift: & $0.4270$ \\
Velocity dispersion: & $1490_{-130}^{+104}$ km s$^{-1}$ \\ 
Dynamical mass ($r{<}r_{\rm 200}$): & $M{=}4.2^{+0.9}_{-1.0}\times10^{15}$ M$_\odot$ \\ \hline 
Within $r{<}r_{\rm 1000}$:\\
X-ray flux (0.5--7 keV): & $(4.10\pm0.05){\times} 10^{-12}$ erg cm$^{-2}$ s$^{-1}$\\
ICM temperature & $(12.4\pm0.6)$ keV\\
X-ray luminosity (0.5--7 keV): & $(3.08\pm0.03){\times} 10^{45}$ erg s$^{-1}$\\
X-ray luminosity (0.1--2.4 keV): & $2.29_{-0.03}^{+0.02}{\times} 10^{45}$ erg s$^{-1}$\\
X-ray luminosity (bolometric): & $3.88_{-0.05}^{+0.04}{\times} 10^{45}$ erg s$^{-1}$\\ \hline
Strong-lensing mass ($r{<}r_{\rm 2500}$): & $(7.28{\times}10^{14})$ M$_\odot$\\
\end{tabular}
\caption{Global properties of MACSJ0553. \label{tab:globprop}}
\end{table}

\subsection{Galaxy distribution}
\label{sec:gals}

\begin{figure}
\hspace*{0mm}\includegraphics[width=0.48\textwidth]{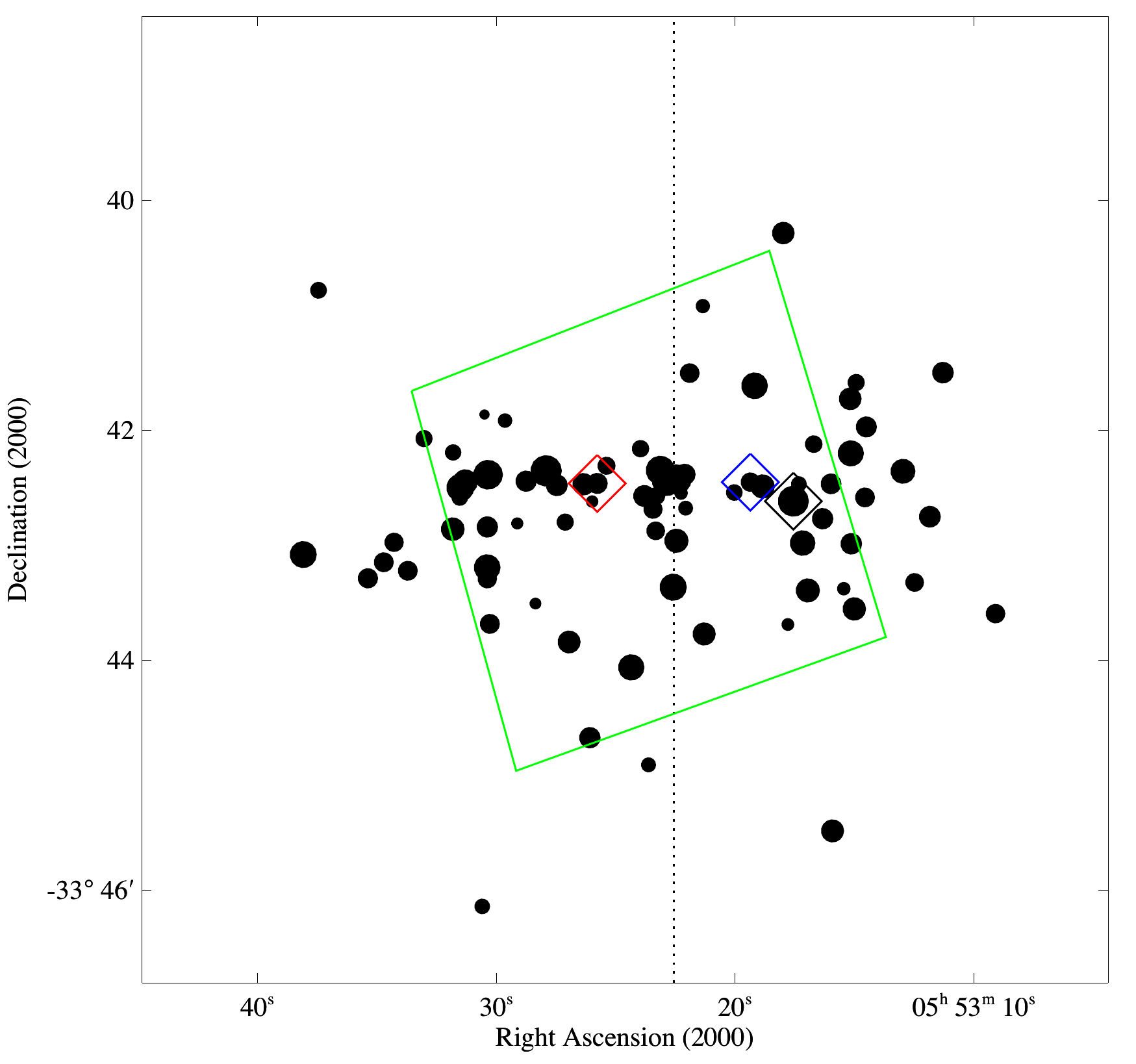}
\caption{Distribution of spectroscopically confirmed cluster members on the sky. Symbol size is proportional to redshift; the two BCGs are marked by red and blue diamonds, respectively. Also highlighted by a diamond is the ``jellyfish" galaxy southwest of the western BCG. The outlines of the ACS field of view are shown in green. The dotted vertical line divides the galaxy distribution at the right-ascension midpoint between the two BCGs. \label{fig:zsky}
}
\end{figure}

\begin{figure}
\hspace*{-5mm}\includegraphics[width=0.52\textwidth]{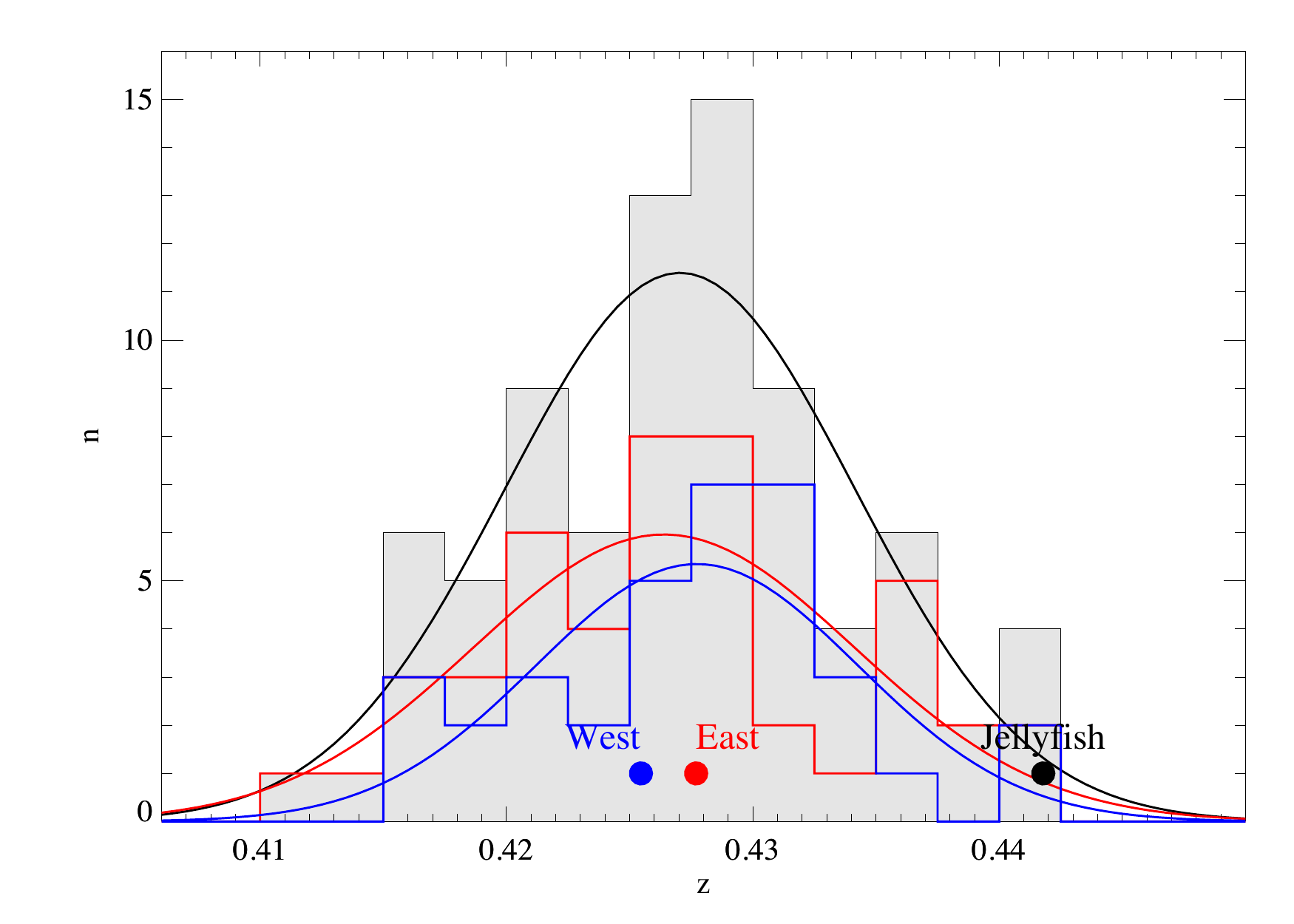}
\caption{Redshift distribution of cluster members. When converted to co-moving velocities, the best-fit Gaussian has a velocity dispersion of 1490 km s$^{-1}$.  The redshift histograms for the eastern and western parts of the distribution (divided by the dotted line of Fig.~\ref{fig:zsky}) are characterised by similarly high velocity dispersions; their offset from each other is not statistically significant. The redshifts of the western and eastern BCG, as well as of the jellyfish galaxy discussed in Section~\ref{sec:jelly}, are marked.   \label{fig:zhist}
}
\end{figure}

We list, in Appendix \ref{sec:app1}, the positions and redshifts of all galaxies successfully targeted in our spectroscopic follow-up observations (Section~\ref{sec:spec} and Table~\ref{tab:obsruns}). Their location on the sky is shown in Fig.~\ref{fig:zsky}. We find no significant substructure along the line of sight, as evidenced by the unimodal redshift distribution shown in Fig.~\ref{fig:zhist}. The overlaid Gaussian, derived from the redshifts of 81 cluster members using the \textsc{Rostat} package \citep{beers90}, is characterised by a velocity dispersion of $\sigma{=}1490_{-130}^{+104}$ km s$^{-1}$. Adopting the $M-\sigma$ scaling relation for dark matter derived by \citet{evrard08}, and assuming that galaxies are unbiased tracers of dark matter, this high value of $\sigma$ implies a very high dynamical mass of $M_{\rm 200}=4.2^{+0.9}_{-1.0}\times10^{15}$ M$_\odot$.

The two BCGs of MACSJ0553, highlighted in Fig.~\ref{fig:hst}, feature redshifts of $z{=}0.4277$ (east) and 0.4225 (west). Along our line of sight, and relative to the eastern BCG, the western BCG thus moves toward us with a velocity of 460 km s$^{-1}$. No significant relative bulk motion is observed though when the galaxy distribution is divided into an eastern and western subset at the right ascension marked by the dotted line in Fig.~\ref{fig:zsky}: the centres of the respective best-fitting Gaussians are located at 0.4264 and 0.4277 (eastern and western subset, respectively) which is insignificant given the uncertainties of 0.001 in the location of the centroid of either Gaussian.

\subsubsection{Foreground group}

As can be gleaned from Table~\ref{tab:galz} (but not from Fig.~\ref{fig:zhist}), the field of MACSJ0553 also contains a likely foreground group at $z{=}0.266$, three members of which have spectroscopic redshifts. Although this system is irrelevant for the physics of the MACSJ0553 merger event, it is noteworthy in as much as its brightest member, highlighted on the top left of Fig.~\ref{fig:hst} was mistaken for a bright cluster member by \citet{harvey15} (see Appendix~\ref{sec:app2} for a discussion of the consequences of this misidentification).

\subsubsection{Jellyfish}
\label{sec:jelly}

\begin{figure}
\hspace*{0mm}\includegraphics[width=0.235\textwidth]{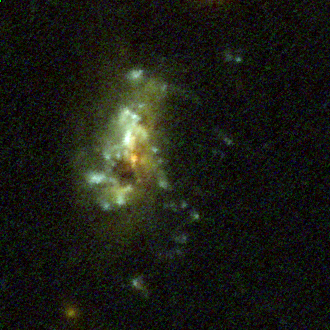}
\hspace*{0.1mm}\includegraphics[width=0.235\textwidth]{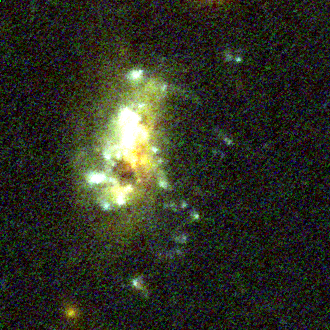}
\caption{A dramatic example of ram-pressure stripping by the western component of MACSJ0553. Shown at two different stretches, this $10{\times}10$ arcsec$^2$ HST/ACS image of MACSJ0553--JFG1 reveals the tell-tale debris trail and thus the direction of motion of an infalling spiral transformed into a ``jellyfish" galaxy.  The BCG of the western component is to the upper left of this image (see Fig.~\ref{fig:hstsl}), suggesting approximately radial infall. \label{fig:jelly}
}
\end{figure}

More relevant in the context of MACSJ0553 is a different, extremely bright galaxy that is a spectroscopically confirmed cluster member. MACSJ0553--JFG1, shown in Fig.~\ref{fig:jelly}, exhibits several characteristics of ongoing ram-pressure stripping, including the suggestive unilateral debris trail of star-forming regions associated with ``jellyfish" galaxies \citep[e.g.,][]{sun07,cortese07,owers12,ebeling14,mcpartland16}.

The location of MACSJ0553--JFG1 is marked in Fig.~\ref{fig:hstsl}; note the excellent alignment of the debris trail with the direction to the centre of the western component of MACSJ0553, 135 kpc away in projection. At a redshift of $z{=}0.4418$, MACSJ0553--JFG1 lies in the high-velocity tail of the redshift distribution shown in Fig.~\ref{fig:zhist} and falls toward the western BCG ($z{=}0.4255$) with a relative line-of-sight velocity of 3400 km s$^{-1}$. The presence of a unilateral debris trail suggests a significant velocity component in the plane of the sky though, implying that the system's true speed must be substantially higher, thus making MACSJ0553--JFG1 one of the fastest infalling galaxies known to date.

\subsection{ICM properties}
\label{sec:icm}
We first determine MACSJ0553's global X-ray characteristics within $r_{\rm 1000}$, the radius within which the overdensity of matter equals 1000 times the critical density\footnote{Since $r_{\rm 1000}$ is computed from the global cluster gas temperature \citep{arnaud02}, this is an iterative process.}.\\

\begin{figure}
\hspace*{-1mm}\includegraphics[width=0.5\textwidth]{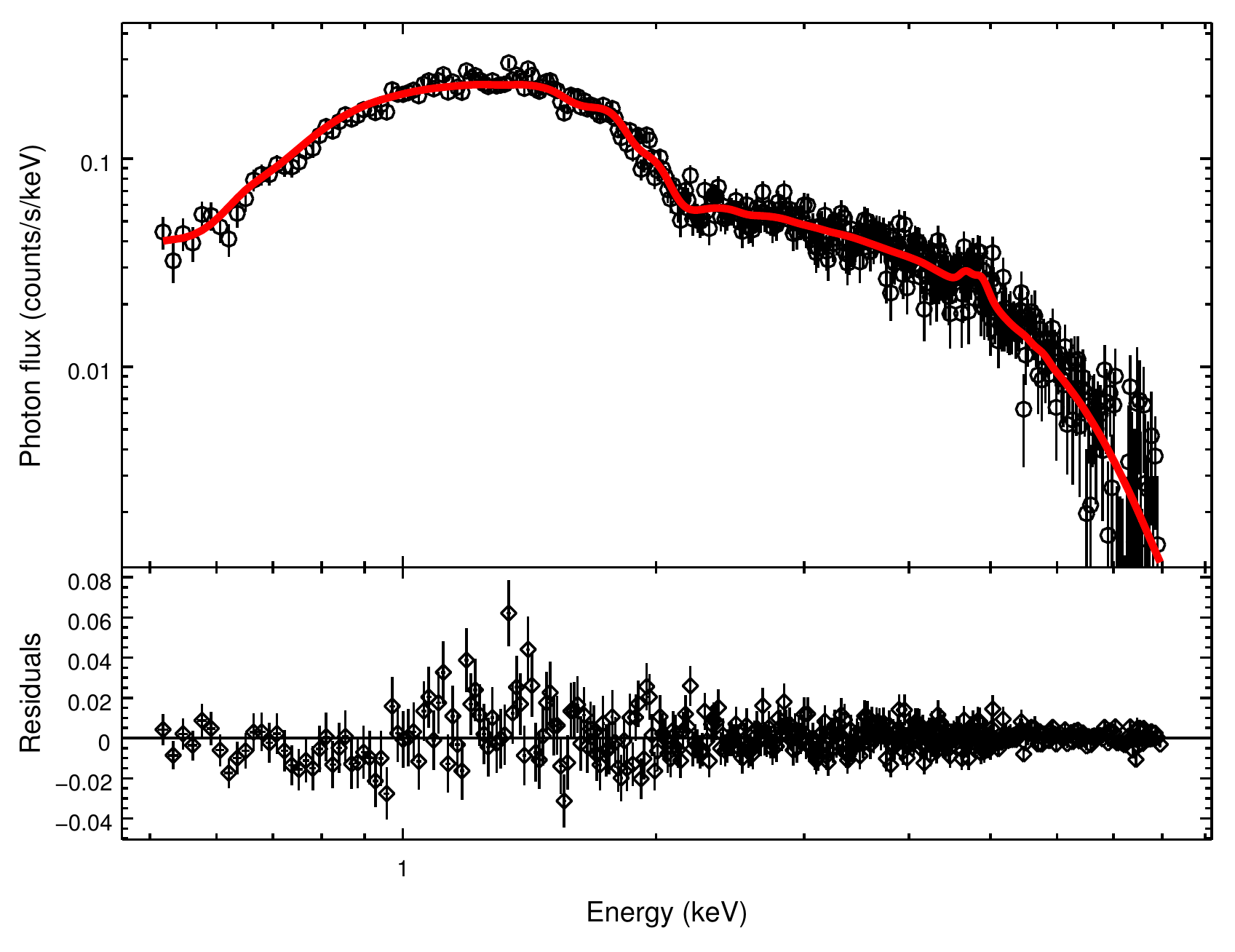}
\caption{X-ray spectrum of MACSJ0553 as seen with Chandra ACIS-I, extracted with $r_{\rm 1000}$. The best-fitting single-temperature plasma model is overlaid in red. The best-fit values for the gas temperature and metal abundance are quoted in Table~\ref{tab:globprop}.   \label{fig:globspec}
}
\end{figure}

For $r_{\rm 1000}{=}1.04$ Mpc (186\arcsec) the best-fitting single-temperature plasma model is described by a redshift of $z{=}0.434^{+0.02}_{-0.03}$, a metal abundance of $Z{=}0.22_{-0.06}^{+0.05}$ Z$_\odot$, an equivalent column density of neutral hydrogen of $n_{\rm H}{=}4.3_{-0.5}^{+0.7}\times 10^{20}$ cm$^{-2}$, and a gas temperature of k$T{=}11.5^{+0.7}_{-0.9}$ keV. Noting that the X-ray redshift is fully consistent with the more precisely determined optical redshift, we freeze $z$ at 0.4270; we also freeze $n_{\rm H}$ at the Galactic value of $3.32\times 10^{20}$ cm$^{-2}$ determined from 21-cm observations \citep{kalberla05} and set $Z{=}0.25$, consistent with our best-fit value and with the findings of \citet{ettori15} for the inner regions of massive, disturbed clusters at $z{\sim}0.4$. Repeating the fit, we find a global value of k$T{=}(12.4\pm0.6)$ keV. The global spectrum and the corresponding best-fitting plasma model are shown in Fig.~\ref{fig:globspec}.

\subsubsection{Spatial ICM distribution}
\label{sec:xsb}

The salient features of our target's appearance in the X-ray regime, evident in Fig.~\ref{fig:cxo}, are:

\begin{itemize}
\item Highly elliptical large-scale emission, along a major axis aligned with the vector connecting the two BCGs;
\item A pronounced X-ray peak at the location of the eastern BCG;
\item No evidence of an X-ray luminous core at the location of the western BCG;
\item A second X-ray peak about 200 kpc east of the eastern BCG that is not associated with any galaxy overdensity;
\item A question-mark-shaped trail of bright, diffuse X-ray emission connecting, and extending beyond, the two X-ray peaks;
\item A sharp drop in X-ray surface brightness to the east of the eastern X-ray peak.
\end{itemize}

\begin{figure}
\hspace*{-2mm}\includegraphics[width=0.5\textwidth]{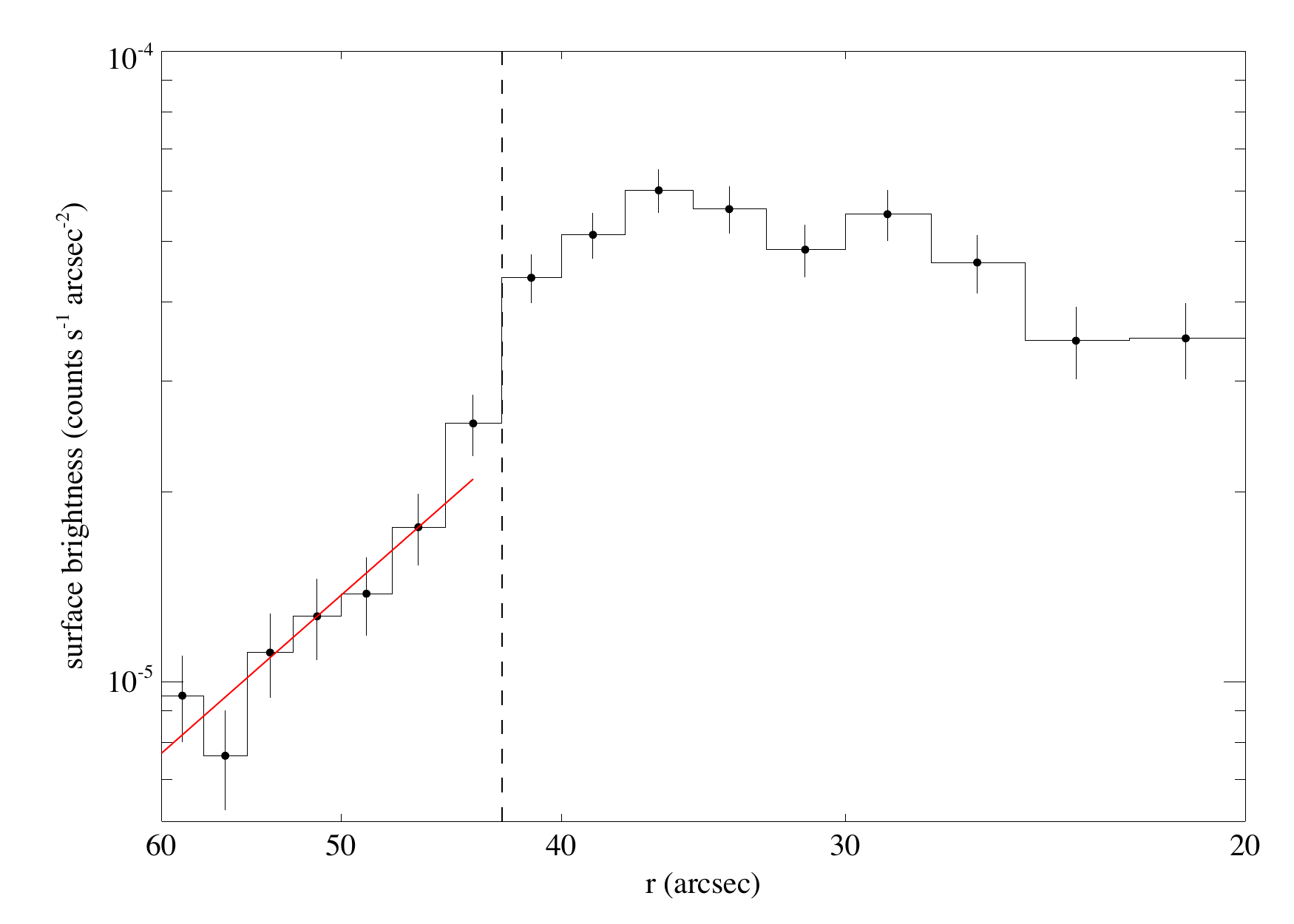}
\hspace*{4mm}\includegraphics[width=0.45\textwidth,clip=true,trim=200 110 250 250]{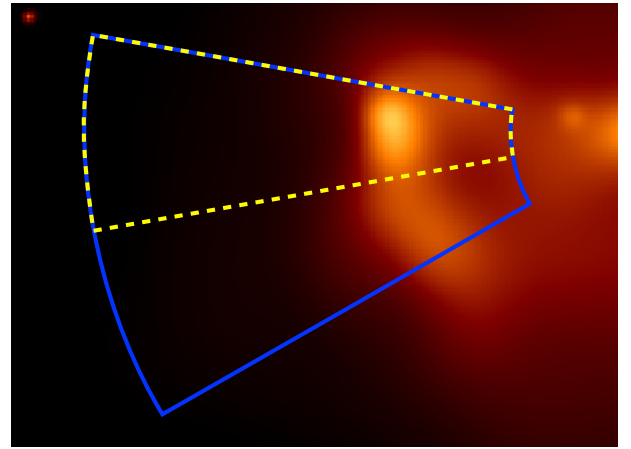}
\hspace*{-2mm}\includegraphics[width=0.5\textwidth]{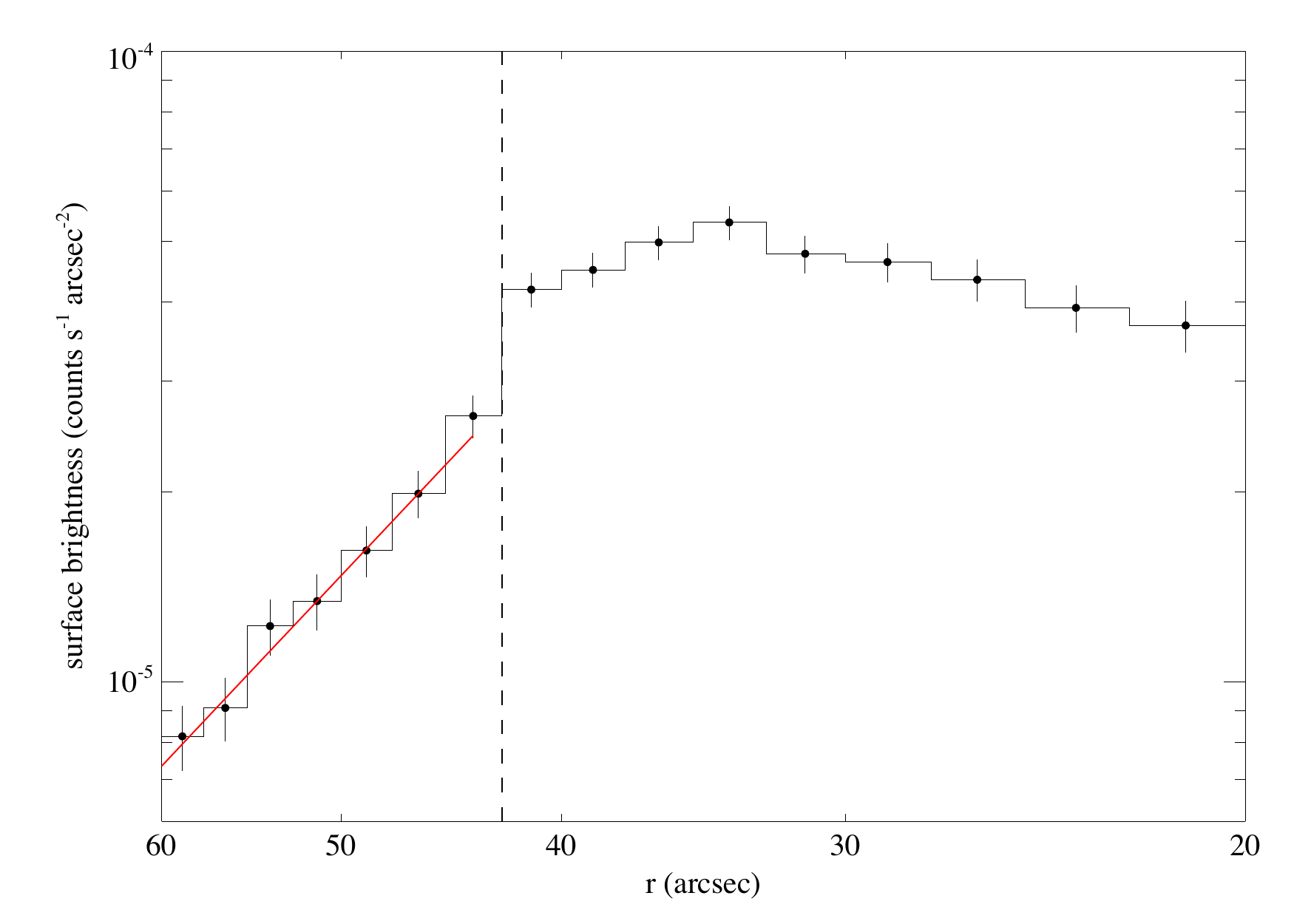}
\caption{Centre: Close-up view of the X-ray emission from MACSJ0553 around the eastern surface brightness discontinuity. The previously mentioned point source is visible in the upper left corner. Top: Surface brightness profile for the upper pie segment, outlined by the dashed yellow line in the central panel; shown on the x axis is the radial distance from the brightest X-ray peak that coincides with the eastern BCG in MACSJ0553. The best-fitting power-law description of the emission in the cluster outskirts is shown in red. Bottom: As top panel, but showing the data extracted from the wider pie segment, outlined in blue in the central panel.  \label{fig:sbp}
}
\end{figure}

We examine the area referred to in the last point above in more detail, in order to assess whether the apparent surface brightness discontinuity is consistent with a shock front or cold front. Fig.~\ref{fig:sbp} (centre) shows a close-up view of the respective part of our X-ray image; the two overlaid sectors, designed to isolate the putative front, extend over a radial distance of 60\arcsec and angles of 20 and 40 degrees, respectively. In both regions the extracted surface brightness profile exhibits a sudden drop at a radial distance from the eastern BCG of approximately 43\arcsec\ (240 kpc). The discontinuity is more pronounced (amounting to a drop by about a factor of 2) within the upper region, which contains the eastern X-ray peak, but is more significant in the larger, 40-degree wide segment (Fig.~\ref{fig:sbp}, top and bottom). 

\begin{figure}
\hspace*{-5mm}\includegraphics[width=0.5\textwidth]{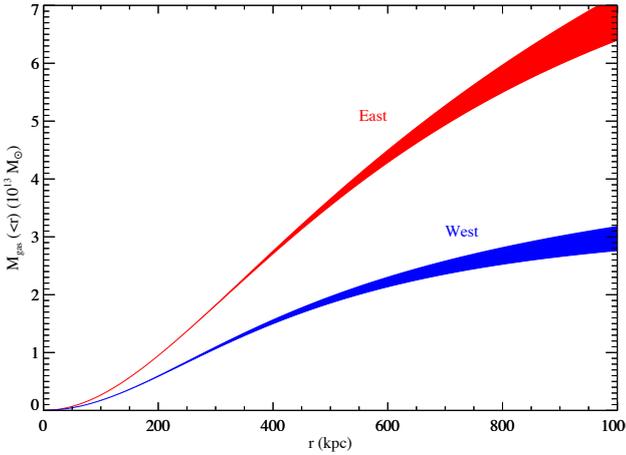}
\caption{Cumulative profiles of the gas mass of the two cluster components based on $\beta$ models. The width of each profile represents the systematic uncertainties introduced by our inability to constrain the parameters characterising the steepening of the emission at large radii and reflect the range of results obtained for all combinations of $r_s=1000, 1500$ kpc and $\epsilon=2, 3$ (see text for details).  \label{fig:mgas}
}
\end{figure}

Motivated by the mentioned complete absence of an X-ray luminous core at the location of the western BCG, we also attempt to quantify the difference in gas density of the two cluster components, following the procedure described in Section~\ref{sec:beta2d}. After elimination of the far-eastern X-ray peak adjacent to the surface-brightness discontinuity discussed above, the remaining emission can be modeled as the superposition of two $\beta$ models. Both components show no significant ellipticity and are aligned with the respective BCG to within about 8 and 12\arcsec (50 and 70 kpc for the eastern and western component respectively), the offsets being mainly toward the south while insignificant within the errors in the east-west direction. We show the resulting cumulative gas mass profiles in Fig.~\ref{fig:mgas}. As expected from even cursory inspection of Fig.~\ref{fig:cxo}, the western component of MACSJ0553 contains significantly less intracluster gas than the eastern one.

\subsubsection{ICM temperature map}
\label{sec:ktmap}

\begin{figure}
\hspace*{-0mm}\includegraphics[width=0.49\textwidth]{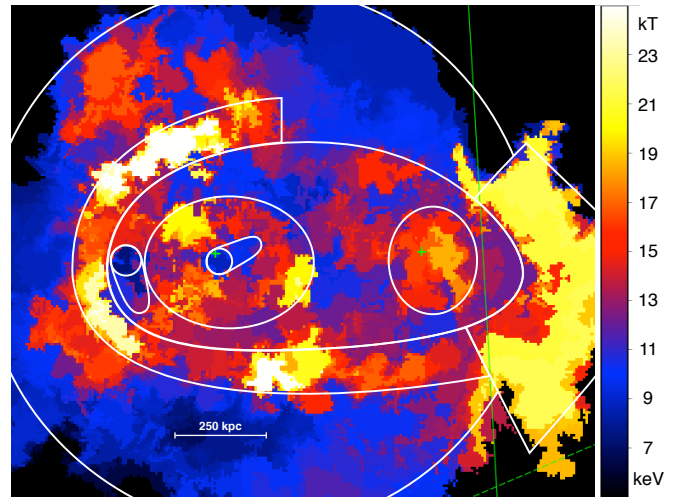}
\caption{ICM temperature trends within MACSJ0553, obtained by median-averaging temperature trend maps for signal-to-noise values of 15, 20, and 25 (see Section~\ref{sec:ttrend}). Delineated in white are the regions defined by us to capture the apparent temperature variations. As in Fig.~\ref{fig:cxo}, the green lines delineate the ACIS-I chip gaps and the green crosses mark the location of the two BCGS.  \label{fig:ktreg}
}
\end{figure}

Following the approach taken by \citet{ma09}, we start our exploration of ICM temperature variations in MACSJ0553 using \textsc{contbin} tessellations for modest SNR values of 15, 20, and 25. Improving upon the strategy of Ma and co-workers, we suppress the geometric noise present in any individual realisation by creating a geometrically averaged ``temperature trend" map as described in Section~\ref{sec:ttrend}. We then median average the trend maps for our three SNR values (15, 20, and 25) to arrive at the map shown in Fig.~\ref{fig:ktreg}.

Inspection of Fig.~\ref{fig:ktreg} reveals several noteworthy features:

\begin{itemize}
\item MACSJ0553 contains one of the \textbf{hottest ICM} observed so far in a galaxy cluster, with ambient temperatures of about 10 keV and extended regions within which the gas temperature is effectively too high to be meaningfully constrained with Chandra (k$T{>}18$ keV);
\item The location of both \textbf{X-ray peaks} coincides with regions containing the coolest gas in the entire cluster (k$T{\sim}8$ keV);
\item Only one of the two X-ray peaks (the western one) aligns with an overdensity of galaxies; it is \textbf{surrounded by much hotter gas} that extends to a radius of 100 to 200 kpc from the cluster core;
\item Although no X-ray peak is observed in this region, the \textbf{ICM at the location of the eastern subcluster} (marked by a very prominent BCG) appears to be much \textbf{hotter} than the ambient gas;
\item An \textbf{arc of extremely hot gas} (k$T{>}20$ keV), extending over some 400 kpc, follows the X-ray surface brightness contours at the eastern end of the cluster, and a similar, but much broader arc of exceptionally hot gas is seen in the far west; both of these features fall into regions of relatively low X-ray surface brightness;
\item Anchored by another region of very hot gas in the south, a \textbf{tentative bridge of heated ICM} connects the two arcs.
\end{itemize}

\begin{figure}
\hspace*{-0mm}\includegraphics[width=0.49\textwidth]{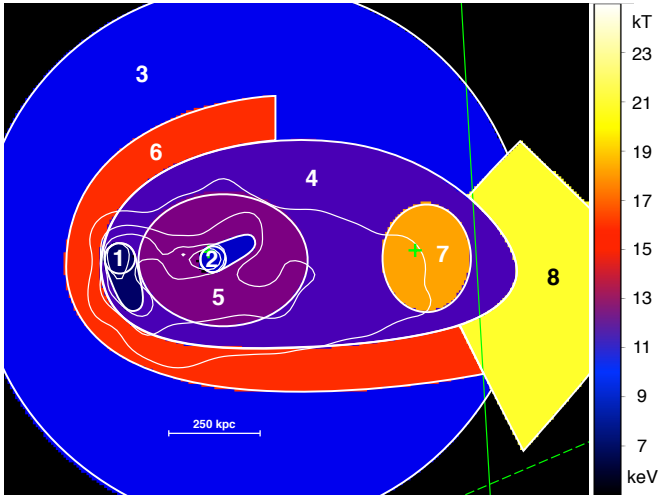}
\caption{ICM temperatures map based on the regions shown in Fig.~\ref{fig:ktreg}. Regions are numbered as in Fig.~\ref{fig:ktranked}. For clarity, we do not label the oval regions associated with the two X-ray peaks and marked with asterisks in Table~\ref{tab:ktmap}.\label{fig:ktmap}
}
\end{figure}

\begin{figure}
\hspace*{-0mm}\includegraphics[width=0.49\textwidth]{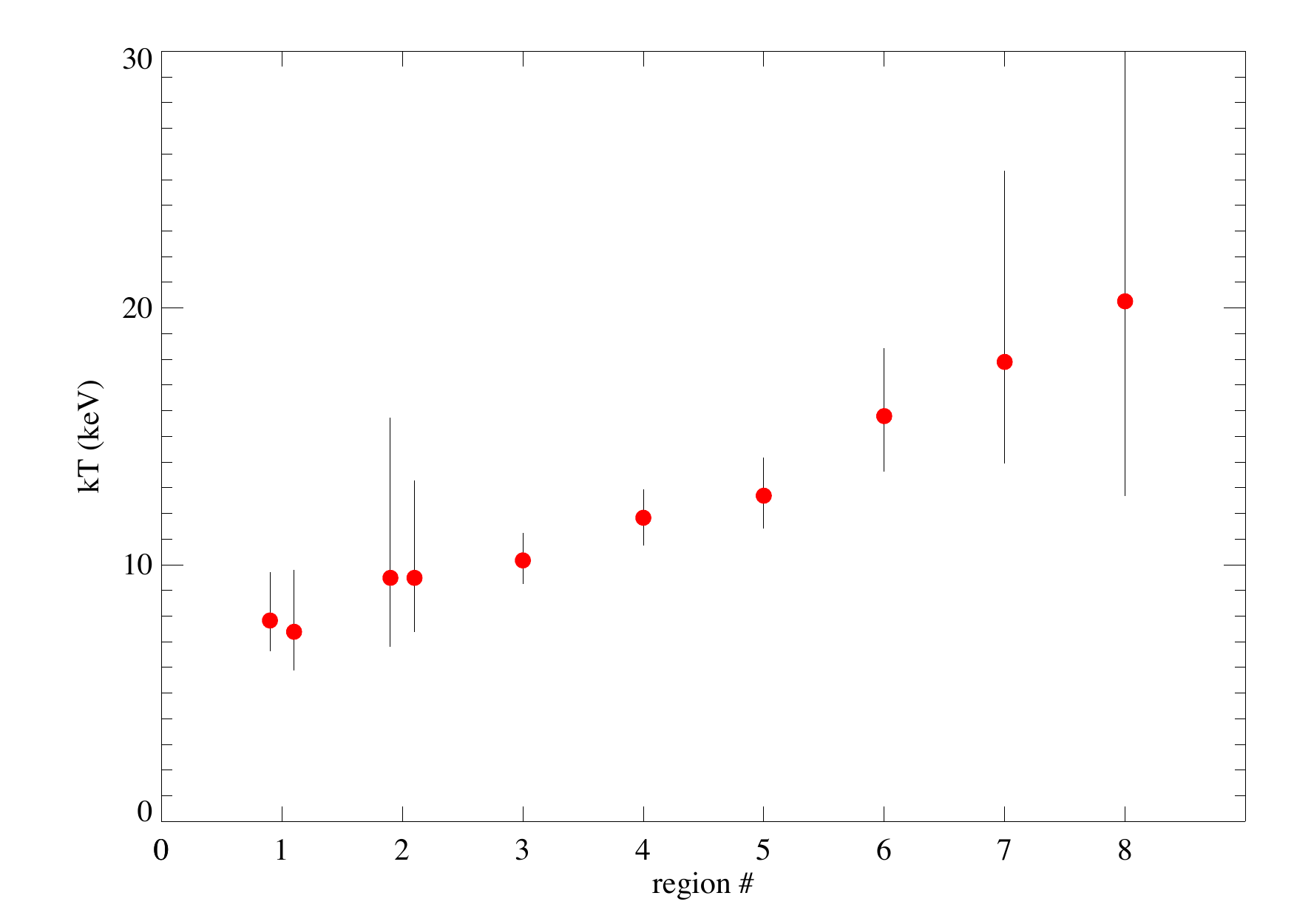}
\caption{ICM temperatures measured within the regions shown in Figs.~\ref{fig:ktreg} and Fig.~\ref{fig:ktmap}, in rank order. Regions can be matched by number with those shown in our coarse temperature map (Fig.~\ref{fig:ktmap}); numerical results are listed in Table~\ref{tab:ktmap}. The dual data points for regions 1 and 2 refer to the circular and oval versions. \label{fig:ktranked}
}
\end{figure}

Since the temperature trend maps used so far are all derived from measurements performed within regions of modest SNR, we now define larger, physically motivated areas designed to isolate the mentioned features, as outlined in Fig.~\ref{fig:ktreg}. The resulting, coarse temperature map for MACSJ0553 is shown in Fig.~\ref{fig:ktmap}; results from the spectral fits are summarised numerically and graphically in Table~\ref{tab:ktmap} and Fig.~\ref{fig:ktranked}, respectively. 

Consistent with the temperature trends discussed above, the coolest gas is found in the vicinity of the eastern X-ray peak; similar temperatures are measured near the western X-ray peak (i.e., the core of the western cluster), and in the outskirts of MACSJ0553. Significantly hotter gas is observed in the eastern and western arcs, but also around the western BCG, where the X-ray surface brightness remains perfectly flat. The sharp increase in temperature of approximately a factor of two between regions 1 and 6 coincides with the surface-brightness drop by about the same factor discussed in Section~\ref{sec:xsb} and shown in Fig.~\ref{fig:sbp}. As a result, the ICM is in approximate pressure equilibrium across this feature, suggesting that it is a cold front \citep{markevitch07}.

Judging from the results of our spectral fits, the ICM in the western region 8 is extremely hot and features temperatures around or above 20 keV that are too high to be meaningfully constrained with Chandra. In addition, the reduced $\chi^2$ values for the four largest regions (\#3, 4, 6, and 8) all exceed unity, suggesting that our model represents an imperfect description of the data. Indeed, the stark temperature differences measured in MACSJ0553 are most likely indicative of multi-phase gas seen in projection. However, attempts to fit two-temperature models fail to yield improved fits at the current photon statistics.

\begin{table}
\begin{tabular}{lcc}
Region \# & k$T$ (keV) & SNR\\ \hline
1              &             $7.4_{-1.5}^{+2.4}$   &      22.1  \\
1$^\ast$   &            $7.8_{-1.2}^{+1.9}$    &      32.4 \\
2              &             $9.5_{-2.7}^{+6.2}$    &     19.1 \\
2$^\ast$   &            $9.5_{-2.1}^{+3.8}$     &      26.1 \\           
3              &            $10.2_{-0.9}^{+1.1}$    &    79.4  \\
4              &            $11.8_{-1.1}^{+1.1}$    &    91.2 \\
5              &            $ 12.7_{-1.3}^{+1.5}$ &  79.3 \\
6              &            $15.8_{-2.2}^{+2.7}$ & 65.5 \\
7              &             $17.9_{-4.0}^{+7.5}$ & 36.3 \\
8              &             $20.3_{-7.6}^{+19.3}$ & 28.5\\
\end{tabular}
\caption{ICM temperatures as measured in regions designed to follow the temperature trends apparent in Fig.~\ref{fig:ktreg}. The same data are shown graphically in Figs.~\ref{fig:ktmap} and \ref{fig:ktranked}. The two regions labeled with an asterisk are the ovals that include, but extend, the circular regions marking the (relatively) cool gas at the two X-ray peaks.\label{tab:ktmap}}
\end{table}

\subsection{Mass distribution: single, binary, or triple?}

\begin{table*}
\begin{tabular}{lccccccc}
Potential & $\Delta\alpha$ & $\Delta\delta$ & $e$ & $\theta$ & r$_{\rm core}$ & r$_{\rm cut}$ & $\sigma$ \\
   & (arcsec) & (arcsec) & & (deg) & (kpc) & (kpc) & (km s$^{-1}$) \\
\hline 
DM1 & $ -5.0^{+  1.3}_{ -1.0}$ & $ -1.4^{+  0.2}_{ -0.2}$ & $ 0.69^{+ 0.03}_{-0.04}$ & $  4.1^{+  0.5}_{ -0.8}$ & $90^{+5}_{-4}$ & $[1000]$ & $1086^{+23}_{-29}$ \\
DM2 & $ 80.0^{+  1.3}_{ -1.0}$ & $ -0.8^{+  0.3}_{ -0.3}$ & $ 0.71^{+ 0.05}_{-0.07}$ & $  9.3^{+  1.2}_{ -1.0}$ & $78^{+10}_{-9}$ & $[1000]$ & $838^{+46}_{-28}$ \\
L$^{*}$ galaxy &  & & & & $[0.15]$ & $47^{+30}_{-12}$ & $204^{+26}_{-13}$\\
\hline
\end{tabular}
\caption{Parameters of our best-fitting model for the total mass distribution of MACSJ0553. Component DM1 and DM2 describe the eastern and western cluster component, respectively. Angular offsets of the centre of each component (second and third column) are measured relative to the position of the eastern BCG.\label{tab:SLresults}}
\end{table*}

\begin{figure*}
\hspace*{-1mm}\includegraphics[width=0.95\textwidth]{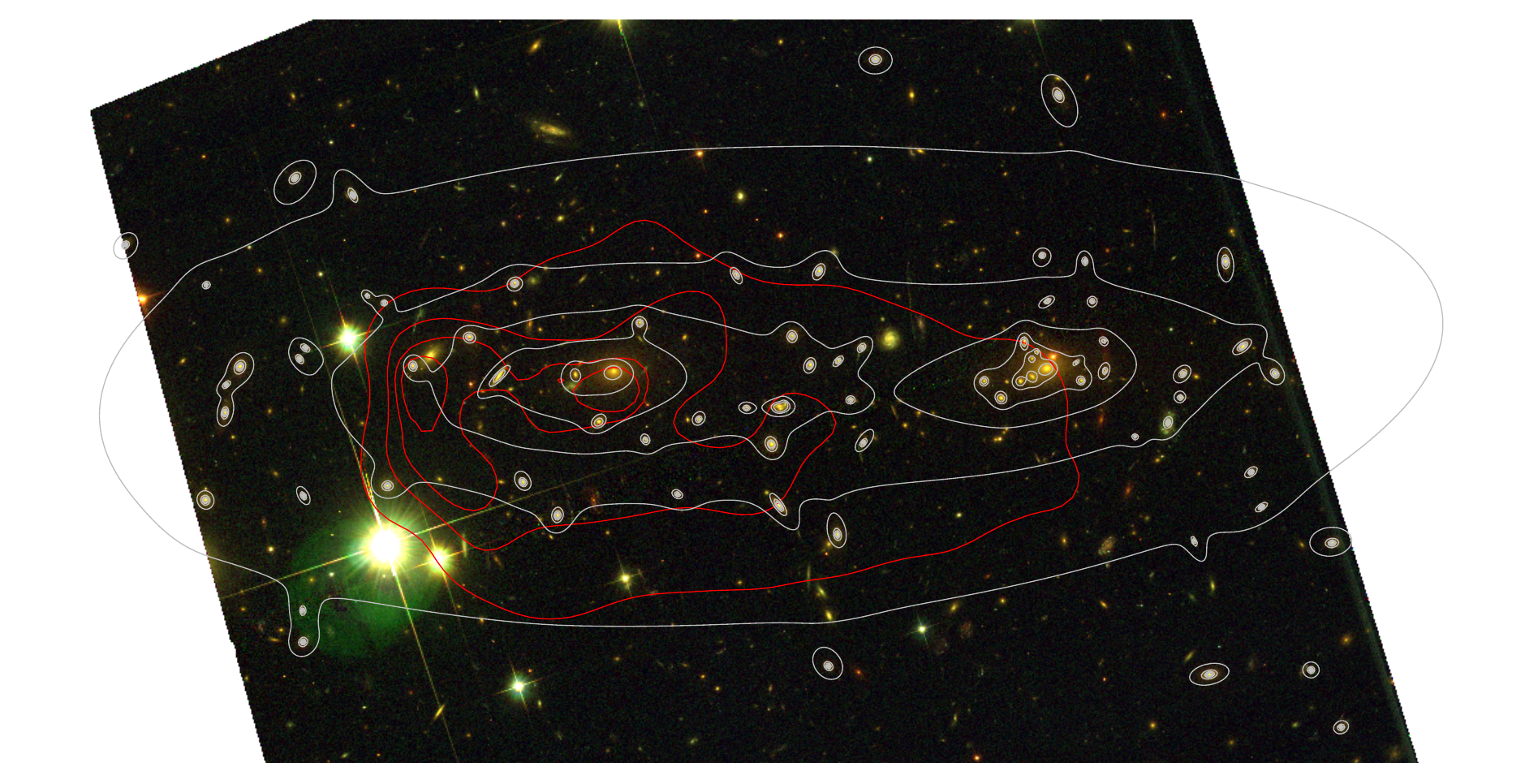}
\caption{Projected surface mass density contours (grey) from our strong-lensing reconstruction overlaid on the HST colour image of MACSJ0553. Red contours show the adaptively smoothed X-ray emission from Fig.~\ref{fig:cxo}. Note the excellent alignment of both components of our mass model with the two BCGs, and the pronounced ellipticity of both mass concentrations.  \label{fig:xraymass}
}
\end{figure*}

\begin{figure}
\hspace*{-1mm}\includegraphics[width=0.48\textwidth]{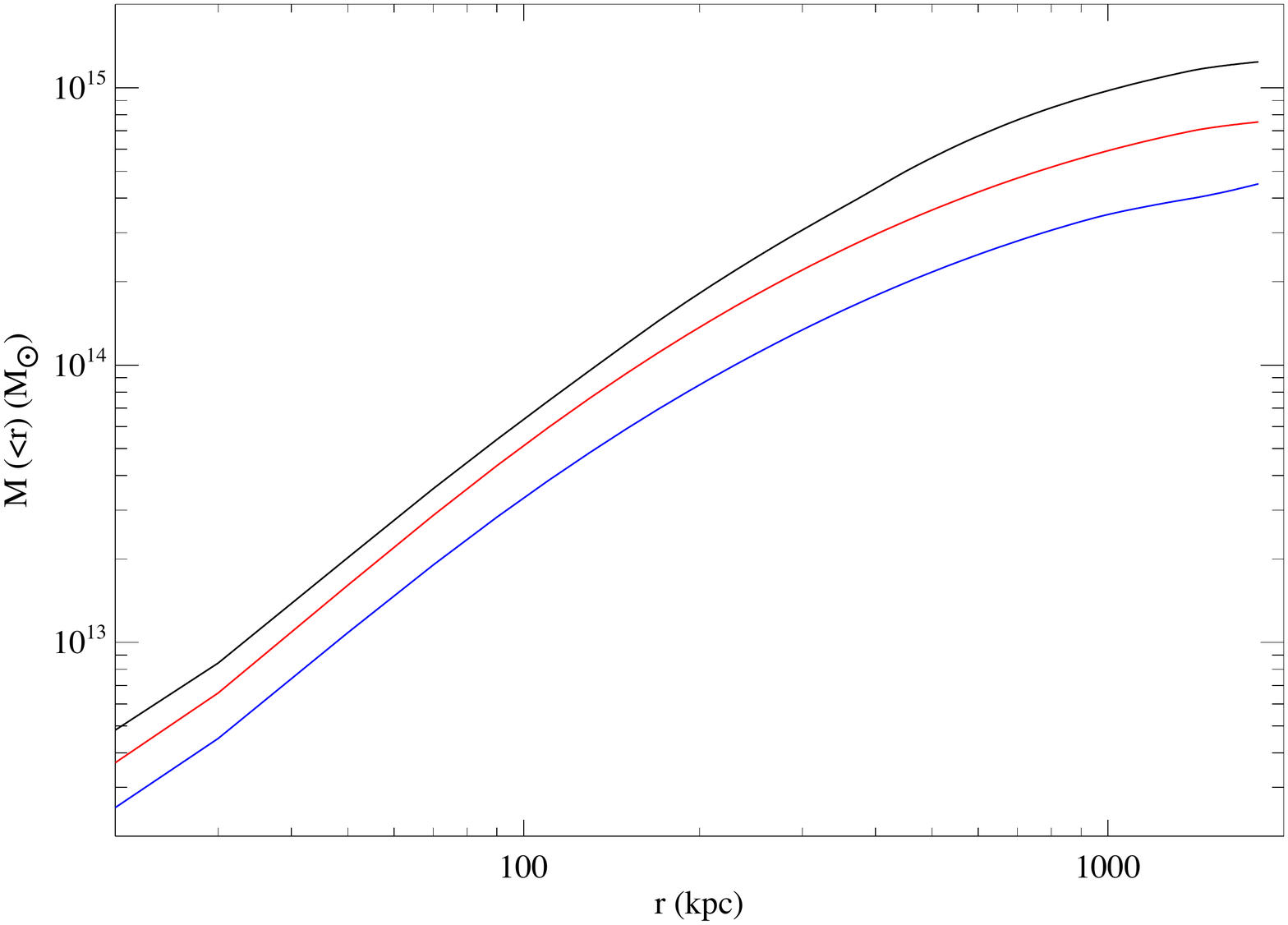}
\caption{Cumulative profile of the total gravitational mass of both cluster components based on our strong-lensing model of MACSJ0553. The red and blue lines show the mass profiles for the eastern and western component, integrated from the location of the respective BCG; the total mass profile, which adopts the position of the eastern BCG as its centre, is shown in black.  \label{fig:mprof}
}
\end{figure}

We tested several assumptions for the large-scale mass components, where we describe the mass distribution as the sum of elliptical mass halos. The large east-west spread of all multiple systems requires a highly elliptical global mass distribution which is not well modeled by a single smooth mass component.  Instead, our baseline model uses two cluster-scale elliptical potentials centred in the vicinity of each of the two BCGs. We use a Gaussian prior for this central location with $\sigma=3\arcsec$ to allow small offsets between the smooth mass distribution and the BCG centre. Such a model reproduces well all multiple systems, with a combined root-mean-square angular offset (rms) of 0.79\arcsec. 

The puzzling presence of a concentration of highly X-ray luminous but relatively cool gas at the eastern edge of MACSJ0553 (see, e.g., Figs.~\ref{fig:cxo} and \ref{fig:ktmap}), in a region that contains hardly any cluster galaxies, prompts us to explore whether our strong-lensing model demands, or at least permits, a third mass component near the location of this eastern-most X-ray peak. We find that the addition of a third smooth mass component, either at the location of said X-ray peak, or in the vicinity of an apparent loose group of cluster members even farther east, fails to attributing significant mass to such a third large-scale halo and also yields no significant improvement of the combined rms. Having also confirmed that our use of Gaussian redshifts priors based on the photometric redshifts instead of a flat redshift prior does not alter the best-fit parameters, we conclude that our extensive lensing constraints are best met by a model consisting of two almost equally massive components.

The resulting best-fit values for the two-component model are reported in Table \ref{tab:SLresults}; the corresponding model predictions for the redshifts of all multiple-image families without spectroscopic redshifts are listed in Table \ref{tab:hstsl}.

Fig.~\ref{fig:xraymass} overlays contours of the X-ray surface brightness distribution from Fig.~\ref{fig:cxo}, as well as contours of the projected mass surface density from our strong-lensing model, onto the HST image of MACSJ0553 and illustrates three key insights regarding the distribution of dark and luminous matter in MACSJ0553:

\begin{itemize}
\item All strong-lensing constraints can be satisfied with a model consisting of two components, the centres of which are well aligned with the two BCGs of MACSJ0553;
\item The mass distribution of both components is highly elliptical;
\item While the centre of the eastern mass component coincides with the western X-ray peak, we have an orphaned X-ray peak to its east (no associated mass or galaxy concentration) and no associated X-ray peak within over 600 kpc of the western component.
\end{itemize}

\subsubsection{Gas mass fraction}
\label{sec:fgas}

\begin{figure}
\hspace*{-1mm}\includegraphics[width=0.48\textwidth]{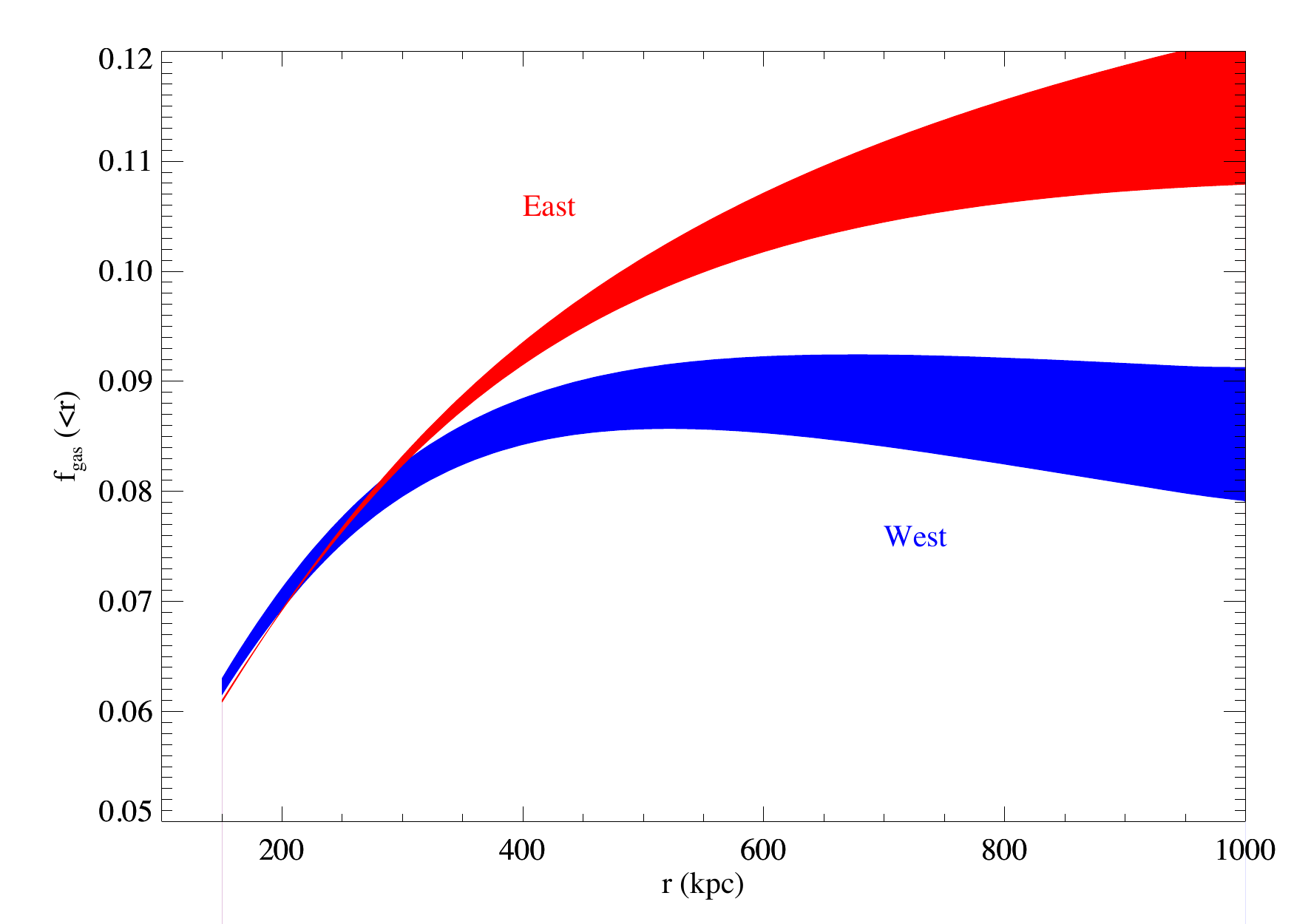}
\caption{Gas mass fraction of the eastern and western component of MACSJ0553 (red and blue curves, respectively), as derived from the profiles shown in Figs.~\ref{fig:mgas} and \ref{fig:mprof}.  As in Fig.~\ref{fig:mgas}, the width of each curve reflects the systematic uncertainties of the simplistic double $\beta$ model used to crudely parameterise the observed X-ray surface brightness distribution of MACSJ0553.\label{fig:gfrac}
}
\end{figure}

\begin{figure}
\hspace*{-1mm}\includegraphics[width=0.48\textwidth]{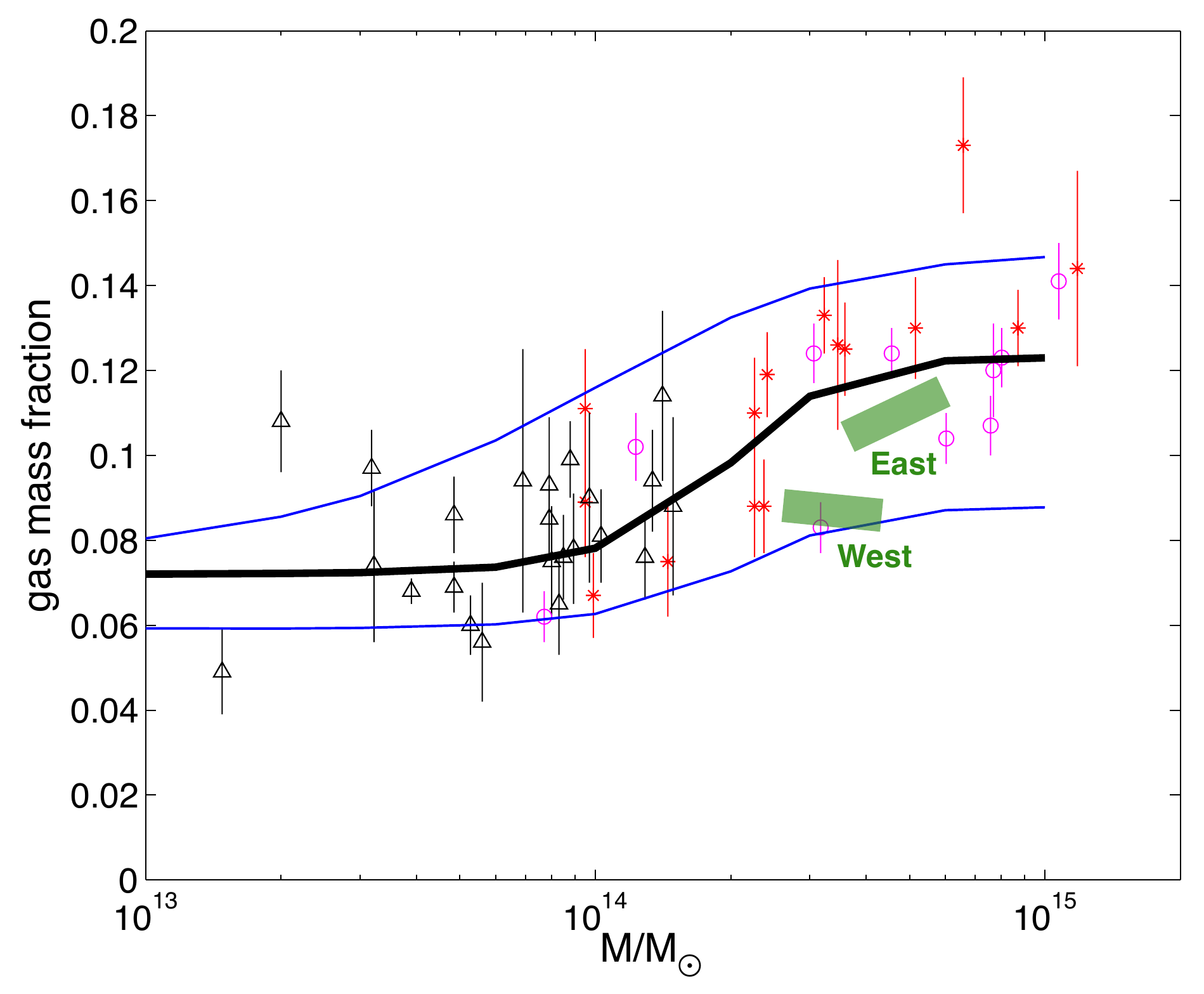}
\caption{Gas mass fractions (at $r<R_{\rm 500}$) as a function of cluster mass. Shown in green are the cumulative gas fractions for the two components of MACSJ0553. Acknowledging the uncertainties in determining $r_{\rm 500}$ for the two subclusters, we show our results for a range of radii between 600 and 1000 kpc  (see Fig.~\ref{fig:gfrac}). The solid black and blue lines mark the model (and associated uncertainties) proposed by \citet{dvorkin15} to describe the data. (Adapted from \citealt{dvorkin15})\label{fig:fgas}
}
\end{figure}

Having established that MACSJ0553 consists of no more than two large-scale mass concentrations, centred upon the two BCGs, we can use the fact that the associated ICM distributions are also reasonably well aligned with the BCGs to constrain the gas fraction of both components by dividing the profiles shown in Figs.~\ref{fig:mgas} and \ref{fig:mprof}. The result is shown in Fig.~\ref{fig:gfrac}. We find the gas fraction of the eastern component of MACSJ0553 to converge toward the universal value of 0.11 \citep{allen08,ettori09}, whereas the western subcluster's much more steeply declining radial X-ray surface brightness profile results in a gas mass fraction that, while not extremely low, drops significantly below that of its eastern neighbour. This difference may still not be noteworthy though, given that the gas mass fraction correlates with cluster mass. To account for this dependency we compare our findings with the compilation of literature values presented in \citet{dvorkin15}. As shown in Fig.~\ref{fig:fgas}, the western component of MACSJ0553 exhibits one of the lowest gas mass fractions measured for clusters in this mass range.

\section{Discussion: merger geometry and history}
\label{sec:geom}

The stark differences between the distribution of dark and luminous matter in MACSJ0553 apparent from Fig.~\ref{fig:xraymass}, as well as the presence of two extended, arc-like regions of very hot, presumably shock-heated gas (Fig.~\ref{fig:ktmap}), represent unambiguous evidence of MACSJ0553 being the result of a massive cluster collision, observed well past pericentre passage. The specifics of our findings, however, pose significant challenges for the interpretation of the geometry and evolution of the merger event. 

\subsection{X-ray / optical / dark-matter alignments}

We first note the excellent agreement between the locations of the eastern BCG, the eastern mass concentration, and the \textit{western} X-ray peak. Although this could be the result of a projection effect, such perfect X-ray / optical alignment is more plausibly accounted for by a tight physical association of all three components. Since the high velocities attained in cluster mergers almost invariably lead to a segregation of collisionless and collisional matter \citep[e.g.,][]{clowe06,bradac08}, we conclude that the eastern component, MACSJ0553-E, must move at negligible speed in the plane of the sky. 

Regarding MACSJ0553-W, the western subcluster, we are left with the puzzling scenario of a massive cluster that is devoid of an associated X-ray peak within over 600 kpc (Figs.~\ref{fig:cxo} and \ref{fig:xraymass}). We stress though that MACSJ0553-W is by no means ``dark": it is clearly discernible as a pronounced galaxy overdensity within a radius of 250 kpc (Fig.~\ref{fig:hst}), and the X-ray luminosity within the same radius is  $6.7\times10^{44}$ erg s$^{-1}$ (0.5--7 keV). MACSJ0553-W lacks, however, any semblance of a dense gaseous core, an unprecedented discovery for a system of such enormous mass ($M{\sim}10^{15}$ M$_\odot$), and indeed features an unusually low gas mass fraction overall (Fig.~\ref{fig:fgas}). 

Rather than trying to find independent explanations for a massive galaxy cluster without a dense gaseous core in the west of MACSJ0553, and for a dense gaseous core without an associated dark-matter and galaxy overdensity in the east, we propose a scenario that naturally and physically links and explains both of these surprising discoveries. In this scenario, the cool core of the ICM in MACSJ0553-W was completely stripped during the initial collision with the even more massive eastern component, and the eastern X-ray peak (which does not coincide with any galaxy or dark matter concentration) marks the remnant of this stripped cool core. If our hypothesis is correct (a more extensive discussion is presented in Section~\ref{sec:merger}), the separation of these two components of over 600 kpc would the largest such offset between dark matter and ICM ever observed in a cluster.

Regarding the X-ray / optical alignment of MACSJ0553-W, the absence of an X-ray peak discussed above prevents us from using any X-ray / optical offset to directly assess the western subcluster's direction of motion and velocity in the plane of the sky. While the latter is somewhat constrained by the good alignment between the location of the BCG and that of a hot region in the ICM temperature map (Fig.~\ref{fig:ktmap}), we also note the clear (if poorly constrained) southerly offset of about 80 kpc of the X-ray emission from the western BCG (Figs.~\ref{fig:cxo} and \ref{fig:xraymass}).

\subsection{Line-of-sight dynamics}

Although the high velocity dispersion of MACSJ0553 may in part be caused by infall, the unimodal redshift distribution shown in Fig.~\ref{fig:zhist} rules out that the two cluster components move at significantly different bulk velocities. Specifically, the redshifts of the BCGs are very similar, their difference amounting to a peculiar velocity of only a few hundred km s$^{-1}$ (Section.~\ref{sec:gals}). We conclude that, in addition to featuring negligible velocities in the plane of the sky, both cluster components also move only slowly with respect to each other along the line of sight.

\begin{figure}
\hspace*{-0mm}\includegraphics[width=0.5\textwidth]{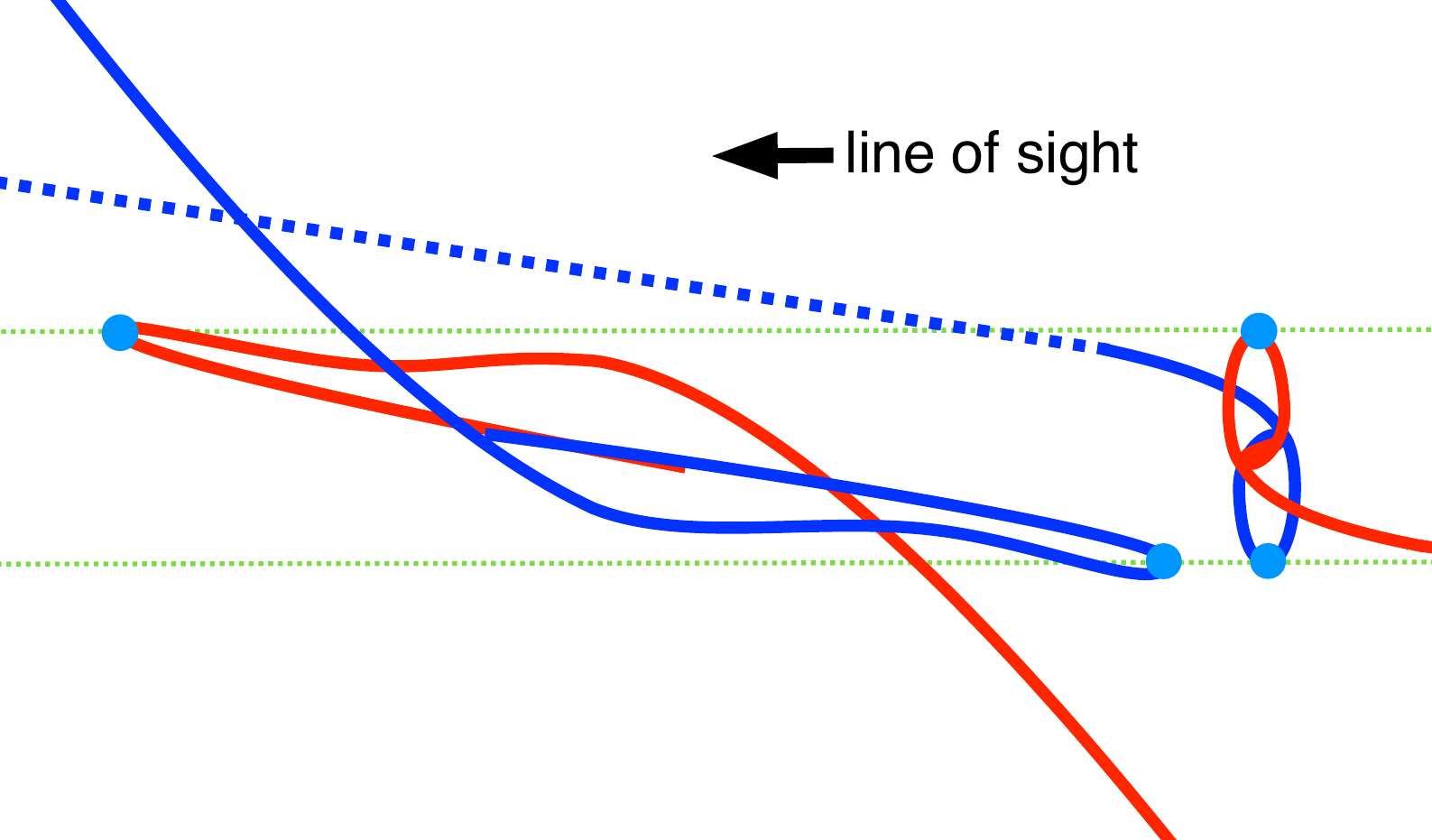}
\caption{Different turnaround scenarios that match the same projected subcluster separation (green dotted lines). Both trajectories (adapted from galaxy-merger tracks by Qi \& Barnes, in preparation) are shown in the centre-of-mass frame for collisions assuming a mass ratio of 1:1 but different impact parameters. The light blue dots mark the locations of the two BCGs at the time of our observation of the system.\label{fig:2turn}
}
\end{figure}

\subsection{Merger scenario}
\label{sec:merger}

We now consider all observational evidence, as presented in the previous sections of this work, to arrive at a self-consistent scenario for the geometry and temporal evolution of the MACSJ0553 merger event.

Although we find two prominent brightest cluster galaxies in MACSJ0553, two well separated X-ray peaks, and two pronounced dark-matter distributions, the relative positions of these features pose serious challenges for the physical interpretation of the merger history.  Considering the absence of X-ray / optical offsets in the eastern cluster core, and the only modest relative line-of-sight velocity of the two subclusters, we propose that the MACSJ0553 merger represents a \textit{two-body collision observed near turnaround}, i.e.,
\[
2r_{\rm t}sin\,i={\rm 450\;kpc}
\]
where $r_{\rm t}$ is the turnaround radius in the centre-of-mass frame, and $i$ is the inclination of the merger axis relative to our line of sight (Fig.~\ref{fig:2turn}).
 The orientation of the merger axis is, in principle, undetermined; a merger axis in the plane of the sky, however, can effectively be ruled out since the resulting maximal separation of the two components of 450 kpc would be unphysically small for systems of such high mass (Fig.~\ref{fig:2turn}). 
 
\begin{figure}
\hspace*{-0mm}\includegraphics[width=0.5\textwidth]{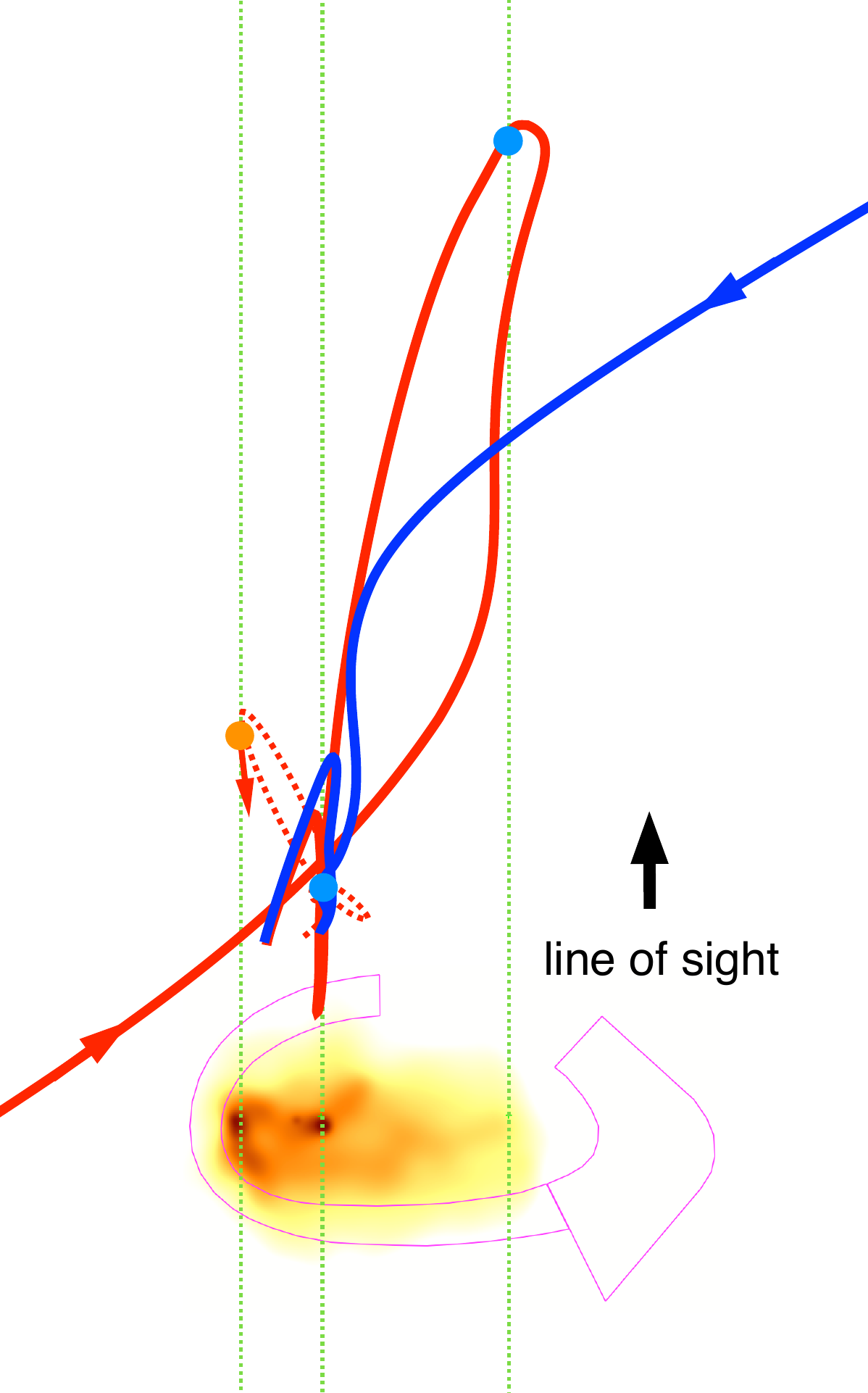}
\caption{Schematic sketch of the MACSJ0553 collision in the centre-of-mass frame. The green, dotted lines mark the projected location of three pertinent features, namely the eastern X-ray peak and the two BCGs. The shown trajectories are adapted from Qi \& Barnes (in preparation) and assume a mass ratio of 5:3 (solid lines) and 1:8 (dashed line). The areas outlined in magenta around the adaptively smoothed X-ray emission shown at the bottom of this plot mark the regions of extremely hot ICM from our temperature map (\ref{fig:ktmap}). \label{fig:sketch}
}
\end{figure}

This scenario does, however, not explain the location of the eastern X-ray peak \textit{outside} the turnaround radius. We therefore put forward the additional proposition that the initial encounter of the two clusters occurred at a very small impact parameter, and that the eastern X-ray peak marks the current position of the cool core of MACSJ0553-W that was completely stripped in the collision and slowed sufficiently to be captured by MACSJ0553-E (Fig.~\ref{fig:sketch}). A schematic trajectory of the captured core, assuming a 1:8 mass ratio for the involved components, is indicated by the dotted line in Fig.~\ref{fig:sketch} and scaled to ensure that the remnant of the stripped cool core is viewed at the observed location\footnote{The real trajectory is bound to be more complicated but accessible to dynamical modeling of what is now a constrained three-body system.}. Note that, in this scenario, the eastern X-ray peak represents the emission from the stripped core moving  along our line of sight and viewed in projection. If we could rotate MACSJ0553 such that its two BCGs fall on top of each other, the stripped core might be similar in appearance to the very extended X-ray tail of a galaxy group undergoing ram-pressure stripping as it falls into A2142, as described by \citet{eckert14}. Although of comparable brightness in X-rays, this highly elongated structure would thus feature a  three-dimensional gas density that is significantly lower than that of the largely undisturbed ICM core of MACSJ0553-W.

Our model explains the hot regions in the temperature map (Fig.~\ref{fig:ktmap}) as follows. Since the peculiar velocities of both MACSJ0553-E and MACSJ0553-W are currently small, no pronounced shock fronts are observed. Shock heating did occur, however, where either cluster component moved at high speed, i.e., primarily close to pericentre passage  (see Fig.~\ref{fig:sketch}). The resulting hot gas has not only had several 100 Myrs to cool and expand; it is also projected on the cooler gas behind and in front of the shocks, thus explaining the lower ICM temperatures measured by us in the central ellipse surrounding the core of MACSJ0553-E. A less obscured view of the shock-heated gas is possible only where it has traveled to larger cluster-centric distances perpendicular to the line of sight, giving rise to the faint but extremely hot ICM observed in the east and west (regions 6 and 8 in Fig.~\ref{fig:ktmap}). Although a merger of two components of almost equal mass would normally lead to a roughly symmetric appearance of the eastern and western expanding shells, MACSJ0553-W leaves a much broader imprint, owing to the loss of its compact core in the encounter with MACSJ0553-E. 

The proposed collision geometry and history finds additional support in the tentative detection of a cold front to the east of what we interpret to be the stripped cool core of MACSJ0553-E (Fig.~\ref{fig:sbp}). Cold fronts are frequently encountered in mergers and created readily during ram-pressure stripping during pericentre passage, consistent with the results of numerical simulations \citep[e.g.,][]{ascasibar06}. We note, however, that cold fronts are subject to Rayleigh-Taylor instability and require either strong gravitational or magnetic support to remain stable \citep{markevitch07}. The lack of any evidence of a dark-matter concentration at the location of the potential stripped cool core thus represents a significant challenge to the cold-front hypothesis.  Part of the hot gas observed along the eastern edge of MACSJ0553 may have been heated by a bow shock that precedes the putative cold front but which, because of the steep inclination of the cool core's velocity vector relative to the plane of the sky, does not manifest itself in a second visible surface brightness discontinuity like the one observed in 1E\,0657$-$56 \citep{markevitch04}. Although cool cores feature the highest gas densities within the ICM and are thus usually more resilient to ram-pressure stripping than the surrounding, more diffuse gas, full stripping of a cool core has been observed  in A520 \citep{wang16}, albeit on much more modest scales. As for the remainder of the ICM of MACSJ0553-W, we stress that the available data (in particular the difference in the gas mass fraction between the two cluster components; Fig.~\ref{fig:fgas}) are consistent with gas loss, but do not allow us to assess how much gas was stripped at pericentre passage; part of the gas that gives rise to the diffuse X-ray emission observed today in the western part of the system may in fact have been (re-)accreted and gravitationally heated after the collision.

Our scenario is consistent also with the diffuse radio emission from MACSJ0553 as mapped by \citet{bonafede12} and shown in Fig.~\ref{fig:radio}. Although radio relics would indeed be expected in the vicinity of the shock fronts, the steep inclination of the merger axis that is a key premise of our merger scenario  will cause them to be projected onto the central radio halo. Polarization measurements that allow halo and relic emission to be disentangled could test this prediction \citep{bonafede12}.

\section{Summary}
\label{sec:summary}

We present a joint X-ray / optical investigation of the extreme cluster merger MACSJ0553 that supersedes the analysis by \citet[][see Appendix~\ref{sec:app2}]{harvey15}. Our strong-lensing reconstruction of the mass distribution unambiguously detects two components, both featuring a very high mass of $M(r{<}r_{1000}){\sim}3{\times}10^{14}$ M$_\odot$ and ${\sim}6{\times}10^{14}$ M$_\odot$ at a separation consistent with that of the two BCGs of the system of 450 kpc. Moderately deep Chandra observations detect no X-ray core within 450 kpc of the western component, but two X-ray peaks in the east, one of them perfectly aligned with the BCG of the eastern subcluster. Our analysis of ICM temperature variations finds two arc-shaped regions of hot gas (k$T{\sim}20$ keV) at the eastern and western edge of the X-ray emission, as well as evidence of cooler gas (k$T{\sim}8$ keV) at the location of either X-ray peak. Moreover, the eastern region of very hot gas is preceded by a surface-brightness discontinuity that we tentatively identify as a possible cold front. Groundbased spectroscopy of cluster members yields a global velocity dispersion of $\sigma{=}1490_{-130}^{+104}$ km s$^{-1}$ and no evidence of velocity substructure along the line of sight. 

We attempt to explain all observational evidence by proposing a merger scenario in which MACSJ0553 represents a two-body collision (mass ratio $\sim$5:3) at small impact parameter, observed at turnaround. Contrary to the earlier assessment of \citet{mann12} and supporting the suggestion made by \citet{bonafede12} based on their analysis of MACSJ0553's radio properties, the merger axis does not lie in the plane of the sky but in fact close to the line of sight. 

We account for the total absence of an X-ray luminous core in the western component by postulating that said core was stripped and captured by the eastern subcluster. Moving at high speed through the inner regions of the eastern cluster, the remnant of this stripped core generates a bow shock and a (potential) cold front seen by us in projection. On larger scales, and again viewed mostly in projection, gas that was shock-heated in the primary collision has expanded to create the arc-shaped regions of extremely hot ICM at both the eastern and western end of the system. The hot gas observed at the far western end is more diffuse, reflecting the profile of the core-less western cluster on its path to turnaround. The ICM of the western component is being gravitationally reheated as it is  (re-)accreted by the  unperturbed deep dark-matter potential around the western BCG.

\section{The future: challenges and opportunities}
\label{sec:future}

If our interpretation is correct at least regarding the fundamentals of the merger geometry, MACSJ0553 presents a rare opportunity to determine reliably the three-dimensional trajectories within a massive cluster merger and hence explore several key science questions, due to a number of fortuitous circumstances:

\begin{itemize}
\item MACSJ0553 is a binary merger, resulting in a single collision plane that renders this target free of the dynamic complexity of multi-body mergers like A520 and A2744 whose three-dimensional merger histories remain almost impenetrable even with extremely deep observations \citep{merten11,wang16};
\item The observation of the merger at turnaround and at an orientation that brings the merger axis close to our line of sight permits a precise measurement of the components' peculiar velocity.
\end{itemize}

Testing our theory regarding the fate of the cool core of the (at present) western subcluster is bound to yield additional insights:

\begin{itemize}
\item The proposed stripping and subsequent capture of the gaseous core of one component by the other requires a very small impact parameter and hence an almost head-on encounter of the densest region of two massive dark-matter haloes. Accurate models of the first core passage of MACSJ0553 would thus not only constrain the collisional and conductive behaviour of hot and cool gas in two massive clusters but also the mass profile in their very core.
\end{itemize}

Although the proposed merger scenario explains much of the observational evidence, it also faces significant challenges. For one, due to the viscous nature of the intracluster medium, any binary cluster collision should result in a gas distribution that peaks between the BCGs (and dark-matter centroids) of the two components, regardless of orientation to our line of sight. We have attempted to explain the violation of this cardinal rule in MACSJ0553 as the consequence of the effective capture of part of one cluster's ICM by the other; however, much better observational data as well as dedicated hydrodynamical simulations of the MACSJ0553 collision will be needed to test the plausibility of the scenario proposed by us here. Both of these (in particular much deeper X-ray observations) will also be critical to solve the puzzle of the survival of the apparent cold front (Section~\ref{sec:xsb} and Fig.~\ref{fig:sbp}) in the absence of gravitational support from an associated dark-matter concentration.

Finally, we note that, if improved observational constraints (from additional spectroscopy of both cluster galaxies and multiple-image systems, as well as deeper X-ray and radio observations) combined with numerical simulations of the collision confirm the scenario schematically outlined in Fig.~\ref{fig:sketch}, the very high central surface mass densities of the two clusters of nearly $3{\times}10^{11}$ M$_\odot$ and the enormous separation of over 600 kpc of the gaseous core of MACSJ0553-W from its galaxy population and dark-matter halo could result in an upper limit to $\sigma/m$ well below 1 cm$^2$ g$^{-1}$, i.e., a value that would be competitive with the constraints set by the Bullet Cluster \citep{randall08}.

\section*{Acknowledgements}
We thank an anonymous referee whose insightful comments greatly improved this paper, to the extent that a co-authorship would be warranted. Support for program GO-12362 was provided by NASA through a grant from the Space Telescope Science Institute, which is operated by the Association of Universities for Research in Astronomy, Inc., under NASA contract NAS 5-26555. Support for this work was also provided by the National Aeronautics and Space Administration through Chandra Award Number GO1-12153X issued by the Chandra X-ray Observatory Center, which is operated by the Smithsonian Astrophysical Observatory for and on behalf of the National Aeronautics Space Administration under contract NAS8-03060. JR acknowledges support from ERC starting grant 336736-CALENDS.

Data presented herein were obtained at the W.M.\ Keck Observatory, which is operated as a scientific partnership among the California Institute of Technology, the University of California, and the National Aeronautics and Space Administration. The observatory was made possible by the generous finical support of the W.M.\ Keck Foundation. Based on observations made with the NASA/ESA Hubble Space Telescope, obtained at the Space Telescope Science Institute, which is operated by the Association of Universities for Research in Astronomy, Inc., under NASA contract NAS 5-26555. These observations are associated with programs GO-12362 and GO-14096.

The authors wish to recognize and acknowledge the very significant cultural role and reverence that the summit of Mauna Kea has always had within the indigenous Hawaiian community.  We are most fortunate to have the opportunity to conduct observations from this mountain.

\appendix

\section{Galaxy spectroscopy}
\label{sec:app1}
We list in Table~\ref{tab:galz} the coordinates and redshifts of all galaxies successfully targeted in our groundbased follow-up observations with optical spectrographs on the Keck twin 10m telescopes.
\begin{table}
\begin{tabular}{lccc}
name & R.A. (J2000) Dec & $z$ & d$z$ \\ \hline
macsg-J055309.1-334336  &   05 53 09.10  -33 43 35.7  &  0.4256  &  0.0003\\
macsg-J055309.7-334455  &   05 53 09.65  -33 44 54.5  &  0.3743  &  0.0002\\
macsg-J055311.3-334130  &   05 53 11.31  -33 41 30.1  &  0.4287  &  0.0001\\
macsg-J055311.9-334245  &   05 53 11.85  -33 42 45.2  &  0.4288  &  0.0009\\
macsg-J055312.5-334319  &   05 53 12.49  -33 43 19.5  &  0.4245  &  0.0003\\
macsg-J055313.0-334222  &   05 53 12.98  -33 42 21.5  &  0.4333  &  0.0001\\
macsg-J055314.5-334158  &   05 53 14.51  -33 41 58.4  &  0.4278  &  0.0003\\
macsg-J055314.6-334235  &   05 53 14.56  -33 42 35.2  &  0.4261  &  0.0002\\
macsg-J055314.9-334135  &   05 53 14.93  -33 41 35.3  &  0.4224  &  0.0001\\
macsg-J055315.0-334333  &   05 53 15.01  -33 43 33.3  &  0.4312  &  0.0009\\
macsg-J055315.1-334259  &   05 53 15.14  -33 42 59.3  &  0.4285  &  0.0004\\
macsg-J055315.2-334212  &   05 53 15.16  -33 42 12.2  &  0.4349  &  0.0000\\
macsg-J055315.2-334144  &   05 53 15.18  -33 41 43.7  &  0.4302  &  0.0004\\
macsg-J055315.4-334323  &   05 53 15.45  -33 43 22.8  &  0.4171  &  0.0010\\
macsg-J055315.9-334529  &   05 53 15.91  -33 45 29.0  &  0.4306  &  0.0003\\
macsg-J055316.0-334228  &   05 53 15.98  -33 42 28.1  &  0.4275  &  0.0010\\
macsg-J055316.3-334246  &   05 53 16.33  -33 42 46.3  &  0.4283  &  0.0004\\
macsg-J055316.7-334207  &   05 53 16.71  -33 42 07.4  &  0.4229  &  0.0005\\
macsg-J055317.0-334324  &   05 53 16.95  -33 43 23.7  &  0.4322  &  0.0001\\
macsg-J055317.2-334259  &   05 53 17.17  -33 42 59.0  &  0.4341  &  0.0007\\
macsg-J055317.3-334228  &   05 53 17.33  -33 42 28.2  &  0.4203  &  0.0002\\
MACSJ0553--JFG1  &   05 53 17.56  -33 42 37.1  &  0.4418  &  0.0001\\
macsg-J055317.8-334341  &   05 53 17.78  -33 43 41.4  &  0.4162  &  0.0005\\
macsg-J055318.0-334017  &   05 53 17.98  -33 40 17.4  &  0.4300  &  0.0001\\
macsg-J055318.5-334300  &   05 53 18.45  -33 42 59.8  &  0.6916  &  0.0005\\
macsg-J055318.8-334229  &   05 53 18.84  -33 42 29.5  &  0.4324  &  0.0007\\
macsg-J055319.2-334137  &   05 53 19.18  -33 41 37.0  &  0.4357  &  0.0001\\
MACSJ0553-W-BCG          &   05 53 19.35  -33 42 27.2  &  0.4255  &  0.0003\\
macsg-J055320.0-334233  &   05 53 20.02  -33 42 32.6  &  0.4216  &  0.0004\\
macsg-J055321.3-334346  &   05 53 21.28  -33 43 46.4  &  0.4307  &  0.0012\\
macsg-J055321.3-334221  &   05 53 21.31  -33 42 21.5  &  0.2657  &  0.0004\\
macsg-J055321.3-334055  &   05 53 21.34  -33 40 55.4  &  0.4183  &  0.0006\\
macsg-J055321.6-334109  &   05 53 21.61  -33 41 08.5  &  0.3242  &  0.0001\\
macsg-J055321.7-334222  &   05 53 21.68  -33 42 21.9  &  0.2670  &  0.0001\\
macsg-J055321.9-334130  &   05 53 21.89  -33 41 30.4  &  0.4259  &  0.0001\\
macsg-J055322.1-334241  &   05 53 22.06  -33 42 40.8  &  0.4191  &  0.0004\\
macsg-J055322.1-334223  &   05 53 22.09  -33 42 23.2  &  0.4283  &  0.0009\\
macsg-J055322.3-334233  &   05 53 22.25  -33 42 32.9  &  0.4172  &  0.0009\\
macsg-J055322.4-334226  &   05 53 22.43  -33 42 25.8  &  0.4404  &  0.0006\\
macsg-J055322.4-334258  &   05 53 22.44  -33 42 57.8  &  0.4322  &  0.0004\\
macsg-J055322.6-334322  &   05 53 22.58  -33 43 22.0  &  0.4365  &  0.0005\\
macsg-J055322.8-334227  &   05 53 22.85  -33 42 26.8  &  0.4397  &  0.0001\\
macsg-J055323.1-334221  &   05 53 23.11  -33 42 21.1  &  0.4394  &  0.0008\\
macsg-J055323.3-334234  &   05 53 23.29  -33 42 34.5  &  0.4235  &  0.0001\\
macsg-J055323.3-334253  &   05 53 23.31  -33 42 52.6  &  0.4246  &  0.0001\\
macsg-J055323.4-334241  &   05 53 23.42  -33 42 41.4  &  0.4251  &  0.0001\\
macsg-J055323.6-334455  &   05 53 23.61  -33 44 54.6  &  0.4191  &  0.0002\\
macsg-J055323.8-334234  &   05 53 23.79  -33 42 34.4  &  0.4291  &  0.0007\\
macsg-J055323.9-334156  &   05 53 23.88  -33 41 55.6  &  0.3348  &  0.0001\\
macsg-J055323.9-334210  &   05 53 23.94  -33 42 09.8  &  0.4226  &  0.0003\\
macsg-J055324.3-334404  &   05 53 24.33  -33 44 03.8  &  0.4352  &  0.0001\\
macsg-J055325.4-334219  &   05 53 25.36  -33 42 18.7  &  0.4236  &  0.0006\\
MACSJ0553-E-BCG           &   05 53 25.75  -33 42 27.9  &  0.4277  &  0.0001\\
macsg-J055326.0-334237  &   05 53 25.97  -33 42 37.3  &  0.4156  &  0.0001\\
macsg-J055326.1-334440  &   05 53 26.06  -33 44 40.5  &  0.4283  &  0.0001\\
macsg-J055326.3-334228  &   05 53 26.32  -33 42 28.3  &  0.4295  &  0.0005\\
macsg-J055326.5-334200  &   05 53 26.54  -33 41 59.5  &  0.3246  &  0.0002\\
macsg-J055326.9-334351  &   05 53 26.93  -33 43 50.6  &  0.4304  &  0.0009\\
macsg-J055327.1-334248  &   05 53 27.09  -33 42 48.1  &  0.4224  &  0.0003\\
macsg-J055327.4-334229  &   05 53 27.44  -33 42 28.9  &  0.4289  &  0.0001\\
macsg-J055327.9-334221  &   05 53 27.89  -33 42 21.3  &  0.4423  &  0.0003\\
\end{tabular}
\caption{Galaxies with spectroscopic redshifts in the field of MACSJ0553. \label{tab:galz}}
\end{table}

\begin{table}
\begin{tabular}{lccc}
name & R.A. (J2000) Dec & $z$ & d$z$ \\ \hline
macsg-J055328.3-334331  &   05 53 28.33  -33 43 30.5  &  0.4147  &  0.0004\\
macsg-J055328.5-334225  &   05 53 28.46  -33 42 25.1  &  0.2658  &  0.0001\\
macsg-J055328.7-334227  &   05 53 28.72  -33 42 26.7  &  0.4280  &  0.0004\\
macsg-J055329.1-334249  &   05 53 29.09  -33 42 48.7  &  0.4152  &  0.0003\\
macsg-J055329.6-334155  &   05 53 29.60  -33 41 55.1  &  0.4186  &  0.0007\\
macsg-J055330.2-334341  &   05 53 30.24  -33 43 41.1  &  0.4262  &  0.0007\\
macsg-J055330.3-334223  &   05 53 30.31  -33 42 23.4  &  0.4400  &  0.0004\\
macsg-J055330.3-334251  &   05 53 30.34  -33 42 50.6  &  0.4282  &  0.0012\\
macsg-J055330.3-334318  &   05 53 30.35  -33 43 17.6  &  0.4254  &  0.0008\\
macsg-J055330.4-334312  &   05 53 30.35  -33 43 11.7  &  0.4356  &  0.0012\\
macsg-J055330.4-334225  &   05 53 30.40  -33 42 25.4  &  0.4162  &  0.0007\\
macsg-J055330.5-334152  &   05 53 30.46  -33 41 52.0  &  0.4124  &  0.0004\\
macsg-J055330.6-334608  &   05 53 30.56  -33 46 08.4  &  0.4201  &  0.0000\\
macsg-J055331.3-334227  &   05 53 31.27  -33 42 27.1  &  0.4337  &  0.0001\\
macsg-J055331.5-334230  &   05 53 31.48  -33 42 30.1  &  0.4373  &  0.0008\\
macsg-J055331.5-334235  &   05 53 31.50  -33 42 35.3  &  0.4215  &  0.0003\\
macsg-J055331.8-334212  &   05 53 31.77  -33 42 11.8  &  0.4212  &  0.0005\\
macsg-J055331.8-334252  &   05 53 31.79  -33 42 51.7  &  0.4314  &  0.0001\\
macsg-J055333.0-334205  &   05 53 32.99  -33 42 04.5  &  0.4224  &  0.0004\\
macsg-J055333.7-334313  &   05 53 33.67  -33 43 13.4  &  0.4261  &  0.0001\\
macsg-J055334.2-334259  &   05 53 34.24  -33 42 58.6  &  0.4253  &  0.0004\\
macsg-J055334.7-334309  &   05 53 34.67  -33 43 09.0  &  0.4262  &  0.0001\\
macsg-J055335.0-334325  &   05 53 34.99  -33 43 25.4  &  0.2187  &  0.0002\\
macsg-J055335.3-334317  &   05 53 35.34  -33 43 17.3  &  0.4266  &  0.0001\\
macsg-J055335.8-334352  &   05 53 35.83  -33 43 51.8  &  0.1923  &  0.0000\\
macsg-J055337.4-334047  &   05 53 37.39  -33 40 47.2  &  0.4218  &  0.0004\\
macsg-J055338.0-334305  &   05 53 38.04  -33 43 04.9  &  0.4362  &  0.0007\\
macsg-J055340.7-334405  &   05 53 40.74  -33 44 05.5  &  0.2048  &  0.0000\\
macsg-J055341.1-334440  &   05 53 41.10  -33 44 39.8  &  0.3313  &  0.0000\\
macsg-J055341.4-334501  &   05 53 41.38  -33 45 01.1  &  1.8800  &  0.0000\\
\end{tabular}
\contcaption{Galaxies with spectroscopic redshifts in the field of MACSJ0553. \label{tab:galz}}
\end{table}

\section{Analysis by Harvey et al.}
\label{sec:app2}

\begin{figure}
\hspace*{-1mm}\includegraphics[width=0.45\textwidth,clip=true,trim=1 0.6 0 0.4]{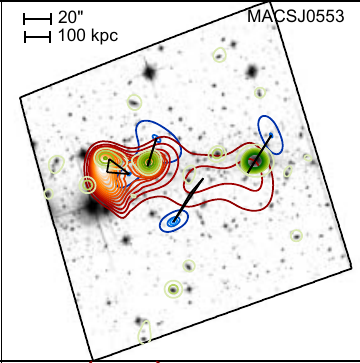}
\caption{Summary of the findings of \citet{harvey15} regarding the distribution of dark and luminous matter in MACSJ0553. Blue contours represent the mass distribution as reconstructed from a weak-lensing analysis; red contours represent the X-ray surface brightness; green contours describe the galaxy light distribution. The data underlying their work are the same as the ones used by us here. (Reproduced from \citet{harvey15})   \label{fig:harvey}
}
\end{figure}

A study aimed at mapping the distribution of gas, galaxies, and dark matter in this system was undertaken before. \citet{harvey15} used the same data for MACSJ0553 (from HST and Chandra) to measure offsets between dark and luminous matter, the former being probed through a weak-lensing analysis of the HST imaging data. Their findings, summarised in Fig.~\ref{fig:harvey}, stand in stark conflict with ours, as their weak-lensing mass reconstruction fails to detect either of the two main mass concentrations in MACSJ0553, and the easternmost peak in their cluster light distribution is based on the erroneous inclusion of the foreground galaxy macsg-J055328.5-334225 ($z{=}0.266$; see Table~\ref{tab:galz} and Fig.~\ref{fig:hst}).

The distribution of dark and luminous matter in MACSJ0553 derived from our analysis thus supersedes the respective results of Harvey and coworkers, which have been called into question also for other clusters in their sample \citep{wittman17}.

\begin{thebibliography}{999}
\bibitem[Allen et al.(2008)]{allen08} Allen S.W., Rapetti D.A., Schmidt R.W., Ebeling H., Morris R.G., Fabian A.C., 2008, MNRAS, 383, 879 
\bibitem[Arnaud et al.(2002)]{arnaud02} Arnaud M., Aghanim N., \& Neumann D.M.\ 2002, \aap, 389, 1 
\bibitem[Ascasibar \& Markevitch(2006)]{ascasibar06} Ascasibar Y., Markevitch M., 2006, ApJ, 650, 102 
\bibitem[Atek et al.(2015)]{atek15} Atek H.\ et al., 2015, \apj, 800,18
\bibitem[Beers et al.(1990)]{beers90} Beers T.C., Flynn K., \& Gebhardt K. 1990, AJ, 100, 32
\bibitem[Bertin \& Arnouts(1996)]{bertin96} Bertin E., \& Arnouts S.\ 1996, \aaps, 117, 393 
\bibitem[Boller et al.(2016)]{boller16} Boller T., Freyberg M.J., Tr{\"u}mper J. et al.\ 2016, \aap, 588, A103 
\bibitem[Bolzonella et al.(2000)]{bolzonella00} Bolzonella M., Miralles J.-M.\ \& Pell{\'o} R.\ 2000, \aap, 363, 476 
\bibitem[Bonafede et al.(2012)]{bonafede12} Bonafede A., Br{\"u}ggen M., van Weeren R., et al.\ 2012, \mnras, 426, 40 
\bibitem[Bowler et al.(2015)]{bowler15} Bowler R.A.A.\ et al., \mnras, 452, 1817
\bibitem[Brada\v{c} et al.(2008)]{bradac08} Brada\v{c} M.\ et al., 2008, \apj, 687, 959 
\bibitem[Bruzual \& Charlot(2003)]{BC03} Bruzual G., Charlot S., 2003, MNRAS, 344, 1000 
\bibitem[Cavaliere \& Fusco-Femiano(1976)]{beta76} Cavaliere A., Fusco-Femiano R., 1976, A\&A, 49, 137 
\bibitem[Clowe et al.(2006)]{clowe06} Clowe D.\ et al., 2006, \apj, 648, L109
\bibitem[Cooper et al.(2012)]{cooper12}Cooper M.C., Newman J.A., Davis M., Finkbeiner D.P., Gerke B.F., 2012, ASCL, 1203.003
\bibitem[Cortese et al.(2007)]{cortese07} Cortese L., Marcillac D., Richard J.\ et al.\ 2007, \mnras, 376, 157 
\bibitem[David, Jones, \& Forman(2012)]{david12} David L.P., Jones C., Forman W., 2012, \apj, 748, 120 
\bibitem[Dvorkin \& Rephaeli(2015)]{dvorkin15} Dvorkin I.\ \& Rephaeli Y., 2015, MNRAS, 450, 896 
\bibitem[Ebeling, Edge \& Henry(2001)]{ebeling01} Ebeling H., Edge A.C., \& Henry J.P., 2001, \apj, 553, 668
\bibitem[Ebeling et al.(2006)]{ebeling06} Ebeling H., White D.A., \& Rangarajan F.V.N.\ 2006, \mnras, 368, 65 
\bibitem[Ebeling et al.(2007)]{ebeling07} Ebeling H., Barrett E., Donovan D., Ma C.-J., Edge A.C., van Speybroeck L., 2007, \apj, 661, L33
\bibitem[Ebeling et al.(2010)]{ebeling10} Ebeling H., Edge A.C., Mantz A., Barrett E., Henry J.P., Ma C.-J, van Speybroeck L., 2010, \mnras, 407, 83
\bibitem[Ebeling et al.(2014)]{ebeling14} Ebeling H., Stephenson L.N.\ \& Edge A.C.\ 2014, \apjl, 781, L40 
\bibitem[Eckert et al.(2014)]{eckert14} Eckert D., Molendi S., Owers M., et al.\ 2014, \aap, 570, A119 
\bibitem[Ettori et al.(2009)]{ettori09} Ettori S., Morandi A., Tozzi P., Balestra I., Borgani S., Rosati P., Lovisari L., Terenziani F., 2009, A\&A, 501, 61 
\bibitem[Ettori et al.(2015)]{ettori15} Ettori S., Baldi A., Balestra I., Gastaldello F., Molendi S., Tozzi P., 2015, A\&A, 578, A46 
\bibitem[Evrard et al.(2008)]{evrard08} Evrard A.E., Bialek J., Busha M., et al.\ 2008, \apj, 672, 122-137 
\bibitem[Ford et al.(1998)]{ford98} Ford H.C., Bartko F., Bely P.Y., et al.\ 1998, \procspie, 3356, 234 
\bibitem[Harvey et al.(2015)]{harvey15} Harvey D., Massey R., Kitching T., Taylor A., Tittley E., 2015, Science, 347, 1462
\bibitem[Hickox \& Markevitch(2006)]{hickox06} Hickox R.C. \& Markevitch M.\ 2006, \apj, 645, 95 
\bibitem[Ho, Ebeling, \& Richard(2012)]{ho12} Ho I.-T., Ebeling H., Richard J., 2012, \mnras, 426, 1992 
\bibitem[Jauzac et al.(2015)]{jauzac15} Jauzac M.\ et al., 2015, \mnras, 452,1437
\bibitem[Jullo et al.(2007)]{jullo07} Jullo E., Kneib J.-P., Limousin M., et al.\ 2007, New Journal of Physics, 9, 447 
\bibitem[Kalberla et al.(2005)]{kalberla05} Kalberla P.M.W., Burton W.B., Hartmann D., et al.\ 2005, \aap, 440, 775 
\bibitem[Kimble et al.(2008)]{kimble08} Kimble R.A., MacKenty J.W., O'Connell R.W., \& Townsend J.A.\ 2008, \procspie, 7010, 70101E 
\bibitem[Kneib et al.(1996)]{kneib96} Kneib J.-P., Ellis R.S., Smail I., Couch W.J.\ \& Sharples R.M.\ 1996, \apj, 471, 643 
\bibitem[Kneib \& Natarajan(2011)]{kneib11} Kneib J.-P.\ \& Natarajan P., 2011, A\&ARv, 19, 47
\bibitem[Kuntz \& Snowden(2000)]{kuntz00} Kuntz K.D.\ \& Snowden S.L.\ 2000, \apj, 543, 195 
\bibitem[Limousin et al.(2012)]{limousin12} Limousin M., Ebeling H., Richard J., et al.\ 2012, \aap, 544, A71 
\bibitem[Lumb et al.(2002)]{lumb02} Lumb, D.H., Warwick R.S., Page M., De Luca A.\ 2002, \aap, 389, 93 
\bibitem[Ma et al.(2009)]{ma09} Ma C.-J., Ebeling H., \& Barrett E.\ 2009, \apjl, 693, L56 
\bibitem[Mann \& Ebeling(2012)]{mann12} Mann A.W.\ \& Ebeling H., 2012, \mnras,420, 2120
\bibitem[Markevitch et al.(2004)]{markevitch04} Markevitch M.\ et al., 2004, \apj, 606, 819
\bibitem[Markevitch \& Vikhlinin(2007)]{markevitch07} Markevitch M., Vikhlinin A., 2007, PhR, 443, 1 
\bibitem[Masters \& Capak(2011)]{masters11}Masters D.\ \& Capak P., 2011, \pasp, 123, 638
\bibitem[McPartland et al.(2016)]{mcpartland16} McPartland C., Ebeling H., Roediger E.\ \& Blumenthal, K.\ 2016, \mnras, 455, 2994 
\bibitem[Mellier et al.(1993)]{mellier93} Mellier Y., Fort B., \& Kneib, J.-P.\ 1993, \apj, 407, 33 
\bibitem[Merten et al.(2011)]{merten11} Merten J., et al.\ 2011, MNRAS, 417, 333 
\bibitem[Newman et al.(2013)]{newman13}Newman J.A., Cooper M.C., Davis M., Faber S.M., Coil A.L., Guhathakurta P.\ et al., 2013, \apjs, 208, 5
\bibitem[Oesch et al.(2015)]{oesch15} Oesch P.A.\ et al., 2015, \apj, 808, 104
\bibitem[Owers et al.(2012)]{owers12} Owers M.S., Couch W.J., Nulsen P.E.J.\ \& Randall S.W.\ 2012, \apjl, 750, L23 
\bibitem[Randall et al.(2008)]{randall08} Randall S.W., Markevitch M., Clowe D., Gonzalez A.H., \& Brada{\v c} M.\ 2008, \apj, 679, 1173-1180 
\bibitem[Richard et al.(2009)]{richard09} Richard J., Pei L., Limousin M., Jullo E., Kneib J.-P.\ 2009, \aap, 498, 37 
\bibitem[Richard et al.(2010)]{richard10} Richard J., Smith G.P., Kneib J.-P., et al.\ 2010, \mnras, 404, 325 
\bibitem[Richard et al.(2014)]{richard14} Richard J.\ et al., 2014, \mnras, 444, 268
\bibitem[Randall et al.(2016)]{randall16} Randall S.W.\ et al., 2016, \apj, 823, 94
\bibitem[Sanders(2006)]{sanders06} Sanders J.S.\ 2006, \mnras, 371, 829 
\bibitem[Sun et al.(2007)]{sun07} Sun M., Donahue M.\ \& Voit G.M.\ 2007, \apj, 671, 190 
\bibitem[Trauger et al.(1994)]{trauger94} Trauger J.T.\ et al., 1994, \apj, 435, L3
\bibitem[Vikhlinin et al.(2006)]{vikhlinin06} Vikhlinin A., Kravtsov A., Forman W., Jones C., Markevitch M., Murray S.S., Van Speybroeck L., 2006, ApJ, 640, 691 
\bibitem[Voges et al.(1999)]{voges99} Voges W.\ et al., 1999, \aap, 349, 389
\bibitem[Wang, Markevitch, \& Giacintucci(2016)]{wang16} Wang Q.H.S., Markevitch M., Giacintucci S., 2016, ApJ, 833, 99 
\bibitem[Wittman, Golovich, \& Dawson(2017)]{wittman17} Wittman D., Golovich N., Dawson W.A., 2017, arXiv, arXiv:1701.05877 

\end{thebibliography}
\end{document}